\documentclass[aip,graphicx,reprint]{revtex4-1}

\usepackage{graphicx}
\usepackage{dcolumn}% Align table columns on decimal point
\usepackage{bm}% bold math

\usepackage[utf8]{inputenc}
\usepackage[T1]{fontenc}
\usepackage{mathptmx}

\usepackage{chemformula}
\usepackage{siunitx}
\usepackage{comment}
\usepackage{caption}
\usepackage{subfig}

\newcommand{\rg}{$R_g$ }

\newcommand{\rh}{$R_H$ }
\newcommand{\ratio}{$R_g/R_H$ }
\newcommand{\mel}{$\mu_\text{el}$ }
\newcommand{\qchEffOut}{\langle q_\text{ch}^\text{eff}\rangle_\text{out}}
\newcommand{\qchOut}{\langle q_\text{ch}\rangle_\text{out}}

\begin{document}

\title{Two-step deswelling in the Volume Phase Transition of thermoresponsive microgels}

\author{Giovanni Del Monte}
\affiliation{\mbox{Department of Physics, Sapienza University of Rome, p.le A. Moro 2 00185 Roma, Italy}}
\affiliation{CNR-ISC, Sapienza University of Rome, p.le A. Moro 2, 00185 Roma, Italy}
\affiliation{\mbox{CL2NS, Fondazione Istituto Italiano di Tecnologia, V.le Regina Elena 291, 00161 Roma, Italy}}
\author{Domenico Truzzolillo}
\email[]{domenico.truzzolillo@umontpellier.fr}
\affiliation{\mbox{Laboratoire Charles Coulomb (L2C), UMR 5221 CNRS-Universit\`e de Montpellier, F-34095 Montpellier, France}}
\author{Fabrizio Camerin}
\affiliation{CNR-ISC, Sapienza University of Rome, p.le A. Moro 2, 00185 Roma, Italy}
\affiliation{\mbox{Department of Physics, Sapienza University of Rome, p.le A. Moro 2 00185 Roma, Italy}}
\author{Andrea Ninarello}
\affiliation{CNR-ISC, Sapienza University of Rome, p.le A. Moro 2, 00185 Roma, Italy}
\affiliation{\mbox{Department of Physics, Sapienza University of Rome, p.le A. Moro 2 00185 Roma, Italy}}
\author{Edouard Chauveau}
\affiliation{\mbox{Laboratoire Charles Coulomb (L2C), UMR 5221 CNRS-Universit\`e de Montpellier, F-34095 Montpellier, France}}
\author{Letizia Tavagnacco}
\affiliation{CNR-ISC, Sapienza University of Rome, p.le A. Moro 2, 00185 Roma, Italy}
\affiliation{\mbox{Department of Physics, Sapienza University of Rome, p.le A. Moro 2 00185 Roma, Italy}}
\author{Nicoletta Gnan}
\affiliation{CNR-ISC, Sapienza University of Rome, p.le A. Moro 2, 00185 Roma, Italy}
\affiliation{\mbox{Department of Physics, Sapienza University of Rome, p.le A. Moro 2 00185 Roma, Italy}}
\author{Lorenzo Rovigatti}
\affiliation{\mbox{Department of Physics, Sapienza University of Rome, p.le A. Moro 2 00185 Roma, Italy}}
\affiliation{CNR-ISC, Sapienza University of Rome, p.le A. Moro 2, 00185 Roma, Italy}
\author{Simona Sennato}
\affiliation{CNR-ISC, Sapienza University of Rome, p.le A. Moro 2, 00185 Roma, Italy}
\affiliation{\mbox{Department of Physics, Sapienza University of Rome, p.le A. Moro 2 00185 Roma, Italy}}
\author{Emanuela Zaccarelli}
\email[]{emanuela.zaccarelli@cnr.it}
\affiliation{CNR-ISC, Sapienza University of Rome, p.le A. Moro 2, 00185 Roma, Italy}
\affiliation{\mbox{Department of Physics, Sapienza University of Rome, p.le A. Moro 2 00185 Roma, Italy}}


\begin{abstract}
Thermoresponsive microgels are one of the most investigated types of soft colloids, thanks to their ability to undergo a Volume Phase Transition (VPT) close to ambient temperature. However, this fundamental phenomenon still lacks a detailed microscopic understanding, particularly regarding the presence and the role of charges in the deswelling process. This is particularly important for the widely used Poly(N-isopropylacrylamide)-based microgels, where the constituent monomers are neutral but charged groups arise due to the initiator molecules used in the synthesis. Here we address this point combining experiments with state-of-the-art  simulations to show that the microgel collapse does not happen in a homogeneous fashion, but through a two-step mechanism, entirely attributable to electrostatic effects. The signature of this phenomenon is the emergence of a minimum in the ratio between gyration and hydrodynamic radii at the VPT. Thanks to simulations of several microgels with different crosslinker concentrations, charge contents and charge distributions, we provide evidence that peripheral charges arising from the synthesis are responsible for this behavior and we further build a universal master-curve able to predict the two-step deswelling. Our results have direct relevance on fundamental soft condensed matter science and on microgel applications ranging from materials to biomedical technologies.
\end{abstract}

\date{This manuscript was compiled on \today}
\doi{\url{https://doi.org/10.1073/pnas.2109560118 }}

\maketitle

Responsive particles have recently captured the interest of scientists working under many diverse fields~\cite{fernandez2011microgel,lee2017stimuli,mihut2017assembling}. Indeed, their ability to adapt to the environmental conditions has enormous advantages for potential applications from biochemistry to nanomedicine~\cite{medeiros2011stimuli,quinn2017glutathione,wagner2018advanced,wang2019assembling}, but also as smart sensors for various analytes~\cite{jia2018order,shin2018synthesis}.
The versatility of these soft objects lies in the manifold routes in which the chemical components can be synthesized and in the transfer of the single-particle properties to the mesoscopic and macroscopic level. 

In particular, most of these responsive particles are macromolecular colloids, whose inner structure relies on a polymeric system which controls the behavior at the colloidal scale. The prototypical example, that is most actively studied in the literature nowadays, is that of microgel particles, i.e. colloidal-scale realizations of a crosslinked polymer network~\cite{fnieves_lyon,synthesis}. In their most elementary version, these microgels are composed by a single monomeric component. Among all possible compounds, Poly(N-isopropylacrylamide) (pNIPAM) is thermoresponsive  and undergoes a solubility transition from good to bad solvent conditions at a temperature $T_c\sim32$\si{\celsius}. For responsive microgels this phenomenon is called Volume Phase Transition (VPT), by which particles are able to reversibly swell and de-swell  across $T_c$. Microgels can be routinely synthesized in a wide range of sizes roughly going from 50 nm to 100 \si{\micro\meter} in diameter, reason for which they are applicable to a variety of purposes and can be investigated with different experimental techniques, from neutron~\cite{fuzzy} and x-ray scattering~\cite{ninarello2019modeling} up to optical methods and microfluidics~\cite{seiffert2010microfluidic}. In addition, their complex internal structure and collective behaviour, involving particle deformation and interpenetration, can nowadays be resolved with single-particle detail thanks to recent advancements in super-resolution microscopy~\cite{conley2017jamming,conley2019relationship,scheffold2020pathways,bergmann2018super}.
The possibility to be studied with these fascinating tools make them also one of the favorite model systems for fundamental science both in bulk suspensions~\cite{synthesis,bergman2018new} and adsorbed at interfaces~\cite{camerin2020microgels,rey2020poly,fernandez2020microgels}.

For all the above reasons, it is legitimate to say that the volume phase transition occurring in pNIPAM microgels is one of the most studied phenomena in soft condensed matter.  Despite the huge amount of experimental and theoretical work on this topic, that is witnessed by the large number of recent reviews~\cite{yunker2014physics,brijitta2019responsive,karg2019nanogels,quesada_review,rovigatti_review,oberdisse2020recent}, there are still fundamental aspects of the VPT that remain poorly understood.
In particular, pNIPAM microgels are often treated as neutral systems, since electrostatic interactions are usually thought not to play an important role in their behaviour, apart from the stabilization against aggregation given to the suspension, especially at high temperatures. However, the typical batch synthesis procedure of pNIPAM microgels usually includes charged compounds, in particular those from the initiators of the polymerisation process.  While their presence may be effectively neglected or screened out by the addition of salt~\cite{ninarello2019modeling}, recent works pointed out a relevant effect of peripheral charges in concentrated suspensions~\cite{scotti2016role}. At present, the influence of these charges on the VPT has not been clarified yet.

One of the reasons for which these aspects have not been investigated so far may be the lack of detailed numerical simulations able to treat a single microgel in a realistic way. To address these gaps, we recently developed a computational method~\cite{gnan2017in} to assemble disordered networks with desired crosslinker concentration and a core-corona structure that closely reproduces experimental behaviour~\cite{ninarello2019modeling,camerin2019microgels}.
After imposing the correct internal structure, we extended our method to properly include the presence of charged monomers with explicit counterions~\cite{del2019numerical}, again validating our results  in the presence of explicit solvent and comparing with available experiments~\cite{del2020charge}. For these reasons, we are now in the condition to carefully assess the effect of initiator charges on the deswelling mechanism of pNIPAM microgels across the VPT. 

By combining simulations with static and dynamic light scattering experiments, here we show that the presence of these charges strongly affects, from a qualitative point of view, the deswelling transition, inducing an inhomogeneous two-step collapse of the microgels with increasing temperature. This is due to the different solvophobicity of pNIPAM and charged groups, respectively, which manifests in the emergence of a minimum in the ratio between the gyration $R_g$ and hydrodynamic $R_H$ radii at the VPT.  First of all, we show that such a minimum is absent for neutral microgels. Second, we analyse in detail the role of the charge distribution throughout the microgel network to assess whether  the initiator groups are preferentially located on the surface of the microgels, as previously hypothesized~\cite{scotti2016role}, but never effectively proven so far.  In order to be able to predict the onset of the two-step deswelling, we further study different combinations of crosslinking ratio, charge content and charge distribution, establishing clear trends in the occurrence of the minimum in $R_g/R_H$. Notably, we  obtain a master-curve for the observed minimum for all simulated microgels when we plot it as a function of the average charge content per chain on the microgel surface, which turns out to be the simplest indicator of the presence of the two-step collapse.

Our work sheds light on the fundamental electrostatic interactions influencing microgel deswelling, which are crucial to correctly describe their assembly and collective behaviour at high temperatures. In addition, it opens up the possibility to {\it a priori} design microgels with desired characteristics and tunable onset of two-step deswelling, which could be exploited to enhance or adjust the potential applications of microgels as smart micro-objects.

\section*{\label{sec:results}Results}
\subsection*{\label{sec:experiments}Experimental results: swelling curves and $R_g/R_H$}
\begin{figure*}[]
	\centering
	\includegraphics[width=\textwidth]{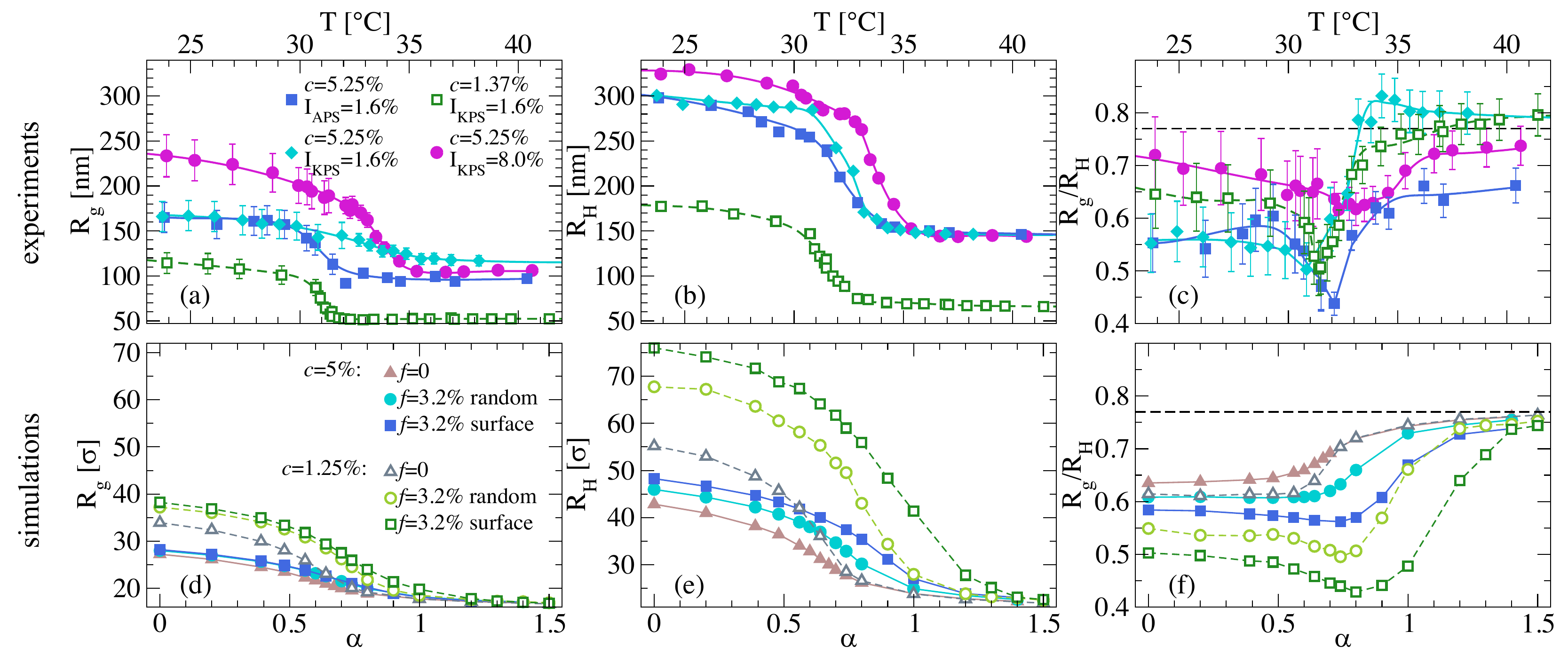}
	\caption{\label{fig1-2} 
		\textbf{Swelling curves and $R_g/R_H$ from experiments and simulations.} (a) Gyration radius $R_g$, (b) hydrodynamic radius $R_H$ and (c) ratio between gyration and hydrodynamic radius $R_g/R_H$ as a function of temperature. Symbols are experimental data, while lines are guides to the eye. Error bars for $R_H$ are smaller than the symbol size;
		(d) $R_g$, (e) $R_H$ and (f) $R_g/R_H$ as a function of the solvophobic parameter $\alpha$ for microgels with $N\sim 42K$ monomers at comparable crosslinker concentrations used in experiments, $c=1.25$\% (open symbols) and $c=5.0$\% (closed symbols). Data refer to neutral microgels ($f=0$, triangles) and to charged microgels with $f=0.032$, corresponding to the initiator amount in experiments, but with two different charge distributions: random (circles) and surface (squares). The horizontal dashed lines in (c,f) indicate the HS value for $R_g/R_H$. }
\end{figure*}
The ratio between gyration and hydrodynamic radii is a relevant quantity in polymer science, widely used as a shape index~\cite{haydukivska2020universal,kok1981relationship}. These two lengths  differently characterize the average polymer distribution inside the macromolecular volume. While $R_g$ is the radius obtained from the standard deviation of the mass distribution around the centre of mass, $R_H$ is an effective size extracted from the self-diffusion coefficient $\bar{D}$ of the particle.
For hard spheres (HS) moving in an ideal continuous solvent, considering a uniform distribution of the mass within the particle, a value of $R_g/R_H \approx 0.77$ is found~\cite{roovers1989hard}.
For pNIPAM microgels obtained through radical polymerisation, this quantity at low $T$ is usually smaller than the HS value~\cite{senff1999temperature}, since particles consist of a denser core with a density profile gradually fading into the corona.  The latter is characterized by the presence of long external chains, the so-called dangling ends, that eventually manifest in an increase of the measured $R_H$, because of their contribution to the drag, but would not affect significantly the value of $R_g$~\cite{nojd2018deswelling}. At high temperatures, instead, where microgels are completely collapsed, the ratio should tend to the homogeneous HS limit. Since these two lengths carry complementary information, it is instructive to compare their behavior for microgels across the VPT, in particular to capture the occurrence of internal inhomogeneities of the macromolecular structure upon collapse. 

We start by showing experimental results for the size variation of microgels in Fig.~\ref{fig1-2}, where both $R_g$ (a) and $R_H$ (b) are reported as a function of temperature $T$ for the different microgel suspensions investigated in this work. 
We consider pNIPAM microgels synthesized with a fixed amount of potassium persulfide (KPS), which brings into the network a negative charge for each radical unit used in the polymerisation reaction. We use a reference sample (i) with a molar fraction of KPS equal to $I_\text{KPS}=1.6\%$ and a crosslinker concentration $c=5.25\%$, in surfactant free conditions, while varying crosslinker content, initiator content and type respectively for three other samples, to systematically assess the effect of each aspect. Namely, we also investigate: 
(ii) $I_\text{KPS}=1.6\%$ and $c=1.37\%$ (with added surfactant, see Methods), primarily to assess the role of crosslinker; (iii) $I_\text{KPS}$=8.0\% and $c=5.25\%$ to evaluate the role of the amount of initiator; (iv) microgels where ammonium persulfide (APS) instead of KPS, is used as initiatior, giving the same charge to the polymer network and differing only on the counterion species (\ch{NH3+} versus \ch{K+}). For the latter case, the amount of initiator and crosslinker are the same as in the reference sample ($I_\text{APS}=1.6\%$, $c=5.25\%$).

We observe in Fig.~\ref{fig1-2}(a) that both $c=5.25\%$ samples with initiator amount of $1.6\%$ have comparable gyration radius of roughly 150 nm. In contrast, the sample with enhanced KPS amount has a much larger gyration radius ($\sim 250$ nm). Concerning the hydrodynamic radius, all three samples with c=5.25\% have a comparable size, close to 300 nm, and a similar swelling ratio roughly equal to 2, as also shown in the Supplementary Information (SI) in Fig.~\ref{fig:normalized}. We notice as well that microgels with c=1.37\% are sensibly smaller, due to the addition of surfactant,  and display a larger swelling ratio, as expected for loosely crosslinked microgels. From the swelling curves, we estimate the VPT temperature $T_c$ using a phenomenological function, as described in the SI. We find that the VPT temperature associated to $R_g$ depends on the amount of initiator, since $T_c\sim31$\si{\celsius} for $I_\text{KPS,APS}=1.6\%$, while it is $\approx2$\si{\celsius} higher for $I_\text{KPS}=8\%$. This result can be rationalized by considering the electric charge brought by initiator molecules. Indeed, in agreement with previous studies on charged microgels~\cite{capriles2008coupled,del2020charge}, a shift of $T_c$ to higher values is found when the amount of charges increases.  In addition, we observe that, for all studied samples, the VPT transition is encountered at a lower temperature, by $\approx 1.0$\si{\celsius}, for $R_g$ with respect to $R_H$. This could be tentatively interpreted in terms of the presence of charged molecules preferentially within the surface of the network~\cite{zhou2012correlation}, an aspect that will be further analyzed later.

Fig.~\ref{fig1-2}(c) shows $R_g/R_H$ as a function of $T$ for all samples. At low temperatures, all studied microgels display a lower ratio with respect to the HS limit, indicating a less homogeneous mass distribution. We observe a higher value for \ratio both for loosely crosslinked microgels (c=1.37\%) and for those carrying a high amount of charges ($I_\text{KPS}=8\%$). The fact that in the former case the particles have a more compact shape, as suggested by the value of \ratio at low $T$, may be caused by the presence of surfactants. Instead, for the latter case, the larger number of chain ends (constituted by the initiator molecules) may result in shorter chains, which, combined with the small crosslinking effect of KPS~\cite{virtanen2016persulfate}, may lead to higher monomers density in the corona. 
However, for large $T$, all microgels become more and more compact with $R_g/R_H$ tending to converge toward the homogeneous HS value. Although experimental uncertainties prevent us from drawing definite conclusions on the specific values and trends observed for the different samples, we clearly identify that all samples display a minimum in $R_g/R_H$ close to the VPT temperature estimated from the hydrodynamic radius (see Table~\ref{tab:1}). In particular, the three microgels with the same initiator amount display a minimum close to $T_c \sim 32$\si{\celsius}, which is instead found at a slightly larger temperature for microgels with $I_\text{KPS}=8\%$. Interestingly, in this case, the minimum is rather shallow, as opposed to those of the other samples where it is more pronounced. 
The presence of a minimum in \ratio has been previously observed in a few experimental studies~\cite{arleth2005volume,sun2005investigation}, including one from some of the present authors~\cite{truzzolillo2018overcharging} for large and loosely crosslinked microgels. However, most of such previous studies reported only a few data points around the VPT, so that the presence of the minimum was hinted, but not totally evident. In the present work, the fine resolution in temperature allows us to unambiguously observe the minimum for all samples.
This can be interpreted in terms of a non-uniform collapse of the microgels, that could be intuitively attributed to the underlying inhomogeneous core-corona structure of the microgels. An alternative explanation could be found in the role played by the initiator charged groups on the deswelling behaviour~\cite{truzzolillo2018overcharging,scotti2016role}.  In particular, arguments in favour of a specific distribution of these charges have been put forward~\cite{zhou2012correlation,huang2017hollow,daly2000temperature}, suggesting that during polymerisation, the charged groups try to minimize their mutual electrostatic repulsion by remaining far from each other, thus being located mostly at the periphery rather than in the core of the microgels. Nonetheless, no direct evidence for the location and the arrangement of these groups has been provided so far. 

\begin{figure*}[]
	\centering
	\includegraphics[width=\textwidth]{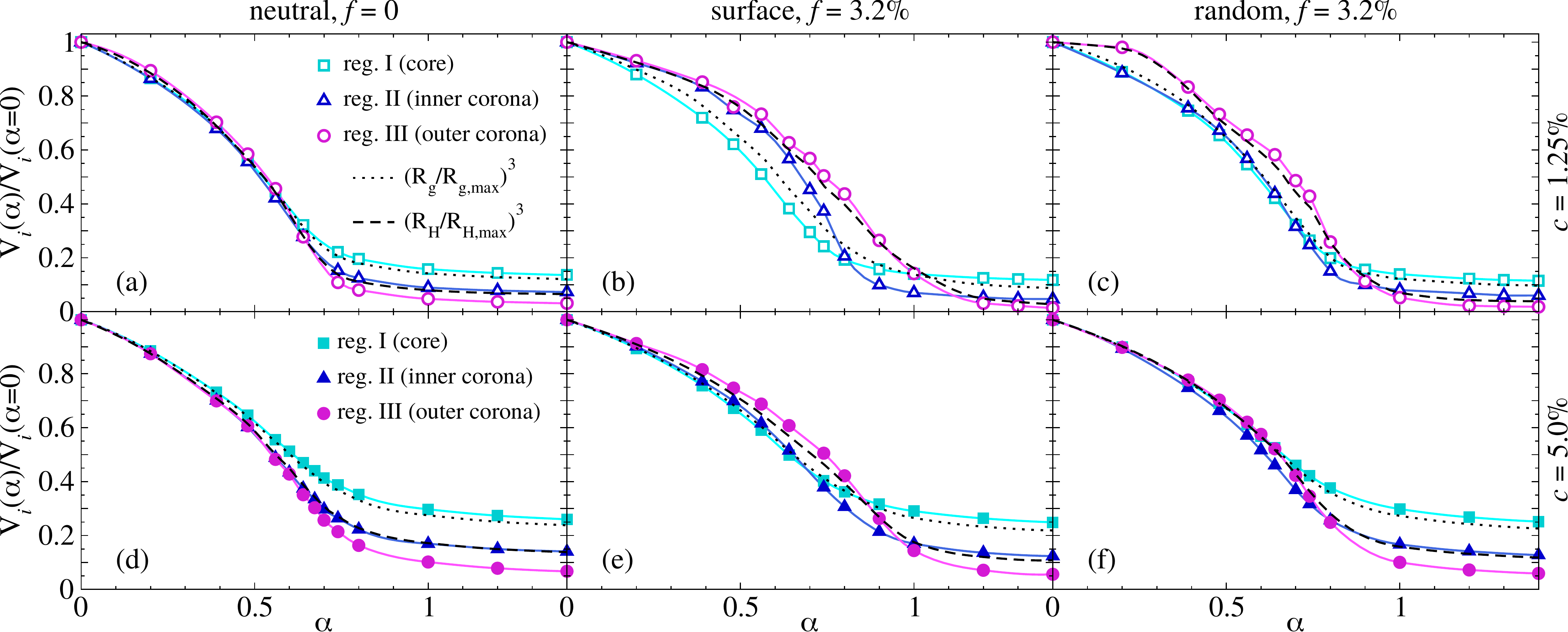}
	\caption{\label{fig:local_swelling} {\bf Local swelling curves.} Normalized volume for the three different regions, defined as $V_{i}(\alpha)/V_{i}(0)$ where $V_i$ refers to the volume of the $i$th region, labeled as $i=$I, II, III for core, inner and outer corona regions, respectively. Top row panels refer to microgels with $c=1.25\%$ and different charge distributions: (a) neutral ($f=0$), (b) surface ($f=3.2\%$)  and (c) random ($f=3.2\%$). Bottom row panels are for microgels with $c=5.0\%$ and different charge distributions: (d) neutral ($f=0$), (e) surface ($f=3.2\%$)  and (f) random ($f=3.2\%$).  Black lines correspond to the volume ratios of spherical regions corresponding to the gyration radius (dotted lines) and to the hydrodynamic radius (dashed lines).
	}
\end{figure*}

\subsection*{\label{sec:simulations}Numerical results: unveiling the role of charges}
To provide a microscopic understanding of this issue, we rely on numerical simulations. We focus on microgels that closely match the experimental ones, by imposing very similar crosslinking contents, namely c=1.25\% and 5\%. We compare the case of perfectly neutral microgels with charged ones, having the same charged monomers fraction ($f=3.2\%$) as the initiator groups in experimental systems. For this specific value of $f$, we also vary the charge distribution by considering a random arrangement of the charges either throughout the network, hereafter named {\it random charge distribution}, or only onto surface chains (see Methods), that we refer to as {\it surface charge distribution}. In this way, we are able to assess how the deswelling behaviour is affected by the charge arrangement upon increasing temperature. 

Our coarse-grained modelling accounts for the presence of a disordered network with a realistic core-corona distribution~\cite{ninarello2019modeling}. In the presence of charges, our treatment explicitly includes the presence of counterions, as these turn out to be crucial to provide a proper description of the local inhomogeneities arising in the network~\cite{del2019numerical}. In addition, charged monomers do not change their affinity to the solvent as we vary the temperature in order to properly capture the experimental situation~\cite{del2020charge}.  A detailed description of the model is provided in Methods.

Numerical results for the radius of gyration $R_g$ are reported in Fig.~\ref{fig1-2}(d) as a function of the solvophobic parameter $\alpha$, which plays the role of an effective temperature in simulations. As expected, the gyration radius is found to be mostly influenced by the crosslinker concentration, with $R_g$ being much larger for $c=1.25\%$  than for $c=5\%$. On the other hand, the presence of charges has a small effect on the gyration radius, which increases only slightly with $f$. Notably, the effect of charge arrangement within the network is almost absent, being $R_g$ very similar for random and surface distributions. 
To estimate numerically $R_H$, we rely on an operative definition that is discussed in Methods and further validated in the SI also by comparing previously published data~\cite{ninarello2019modeling} (see Figs.~\ref{fig:rh_versions},\ref{jerome}).
The resulting $R_H$ are reported in Fig.~\ref{fig1-2}(e) as a function of the solvophobic parameter $\alpha$, displaying a behaviour that looks rather similar to that of $R_g$. An important difference, however, arises when comparing the two different charge distributions. Indeed, we find a significant increase of $R_H$ in the presence of a surface charge distribution compared to the case with random charges due to the accumulation of charges in the corona of the microgel which keeps the latter much more expanded.

\begin{figure*}[]
	\centering
	\includegraphics[width=0.85\textwidth]{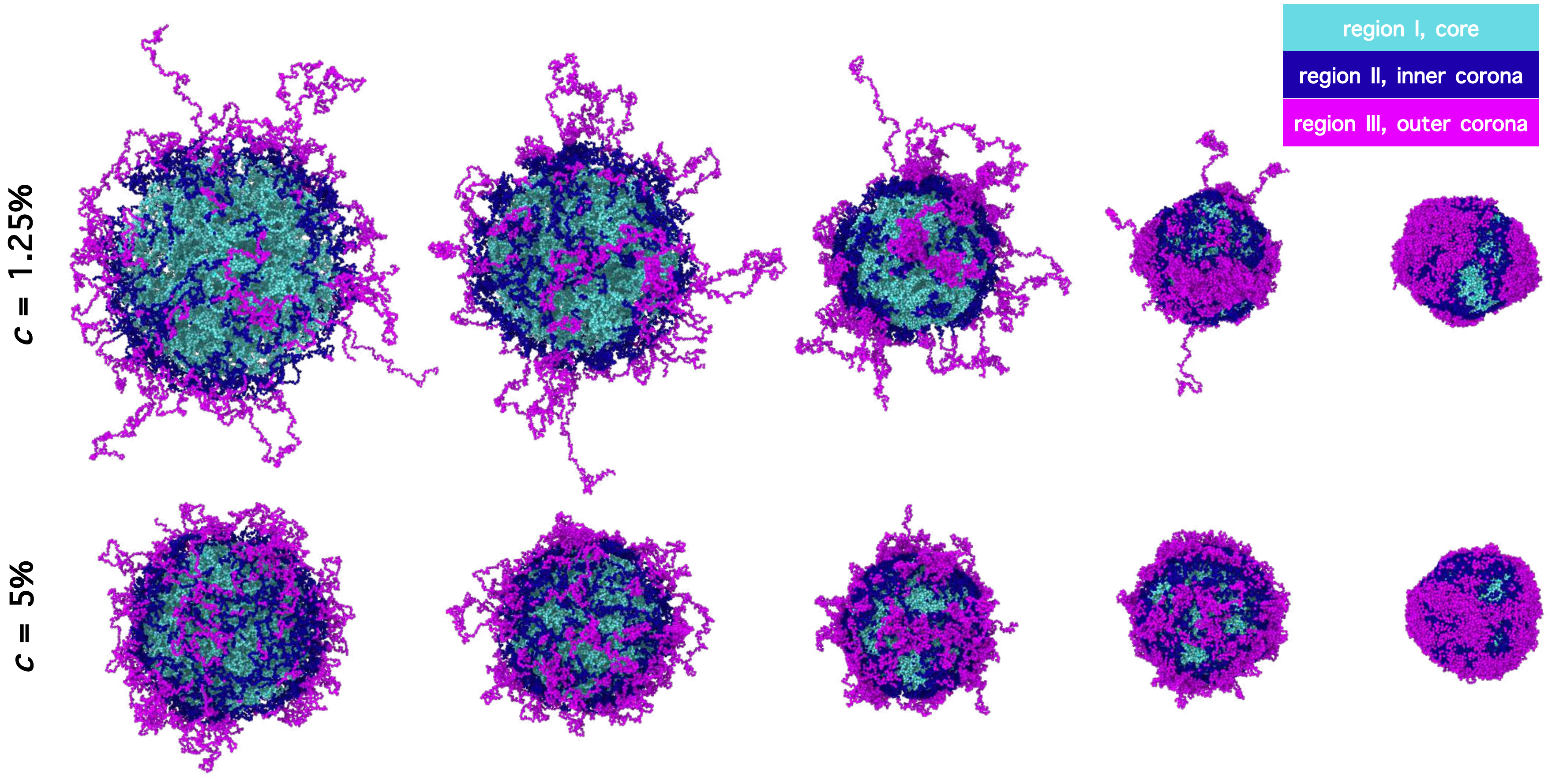}
	\caption{\label{fig:snaps} 
		\textbf{Snapshots of randomly charged microgels across the VPT illustrating the local swelling} for $f=3.2\%$ and two values of  $c$ for different $\alpha$ from the swollen to the collapsed state. From left to right, for $c=1.25\%$, $\alpha=0, 0.48, 0.74, 0.90, 1.20$, while for $c=5\%$, $\alpha=0, 0.48, 0.74, 0.80, 1.20$. Monomers are colored according to the region they belong to: cyan indicate the core region, blue the inner corona region and purple the outer corona region.
		All snapshots refer to equilibrium states. 
	}
\end{figure*}

The calculated ratio $R_g/R_H$ is shown in Fig.~\ref{fig1-2}(f) as a function of $\alpha$. First of all, we notice that our numerical definition of $R_H$ allows us to obtain estimates for the ratio $R_g/R_H$ that are, within statistical uncertainties, in reasonable agreement with experimental ones and that correctly tend to the HS value at large $\alpha$ in all cases. While we observe the emergence of a minimum in \ratio under specific conditions of charge content $f$ and charge distribution, it is important to highlight that such a minimum is completely absent for perfectly neutral microgels, independently of the value of $c$. This result allows us to unambiguously exclude that the appearance of the minimum stems from the underlying core-corona topology of the microgels. Rather, it turns out to be related to the presence of the initiator charges, which therefore plays an explicit role in the VPT of pNIPAM microgels.  Such a role, not often recognized in the literature, appears to be important to properly capture the internal modifications of the microgels upon collapse, as also confirmed by electrophoretic mobility measurements across the VPT for a subset of the samples investigated in this work (see Fig.~\ref{fig:mobility})~\cite{truzzolillo2018overcharging,sennato2021double}.

By examining in more detail the specific features of the microgel for which a minimum in \ratio is present, 
we find that, for both studied values of $c$, microgels with charges arranged on the surface display a very pronounced minimum that occurs roughly at the VPT in agreement with experimental findings. Instead, for randomly located charges, the presence of a minimum is evident only for c=1.25\%, while it is absent for  $c=5.0\%$. Since a minimum is observed for all samples in experiments, this suggests that initiator charges are not randomly spread throughout the network, but rather they are preferentially located close to the microgel surface.

\subsection*{Microscopic origin of the two-step collapse}
In order to unveil the microscopic mechanism that gives rise to the presence of a minimum in $R_g/R_H$, we analyse the microgels by selecting different regions within their volume and calculating their relative swelling in a so-called ``local swelling'' approach. We define the internal regions by fixing the number of particles within each of them on the basis of their distance from the centre of mass, so that, upon varying $\alpha$, we monitor the true volume variation of a given portion of the microgel. The three regions are defined as follows: (region I) the \textit{core} region roughly having the same volume as a sphere of radius equal to the gyration radius; (region II) the \textit{inner corona} region, corresponding to an intermediate shell, whose volume ---together with that of the core--- roughly coincides with that delimited by the hydrodynamic radius, and (region III) the \textit{outer corona} shell. 
More details on the choices of the three regions are given in Methods and in the SI (see Fig.~\ref{fig:local-profiles}).

\begin{figure*}[]
	\centering
	\includegraphics[width=\textwidth]{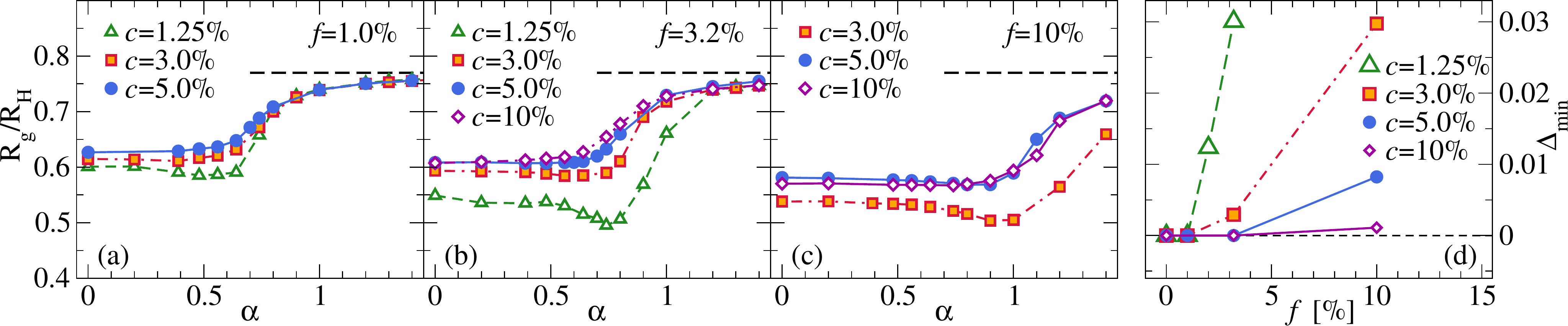}
	\caption{\label{fig:random} 
		\textbf{Ratio \ratio for randomly-charged microgels} as a function of the solvophobic parameter $\alpha$ for (a) $f=1.0\%$, (b) $f=3.2\%$ and (c) $f=10\%$ and different values of $c$. Dashed lines indicate the HS value; (d) Depth  $\Delta_\text{min}$  of the minimum in $R_g/R_H$ as a function of $f$ for different $c$. The corresponding swelling curves are reported in Fig.~\ref{fig:rg-rh-random}.
	}
\end{figure*}

We start by discussing the behavior of neutral microgels. For both values of the examined crosslinked concentrations, reported in Fig.~\ref{fig:local_swelling}(a) and (d), we observe that the deswelling transition, monitored by the volume change of each region, occurs roughly at the same value of the solvophobic parameter for all regions.
These results indicate that a purely neutral microgel deswells in a homogeneous fashion, independently on the internal structure and thus on its intrinsic core-corona topology. This is due to the fact that all monomers experience the same affinity to the solvent, which induces a simultaneous collapse of the whole particle. No difference is observed between the two values of $c$, except for a more pronounced swelling ratio for the less crosslinked microgels, as expected.

We now turn to the surface-charged microgel with $f=3.2\%$, whose local swelling behaviour is reported in Fig.~\ref{fig:local_swelling}(b) and (e).  In this case, it is evident that the collapse of the three regions does not occur simultaneously, with the core (region I) showing a deswelling transition which takes place at a sensibly lower value of $\alpha$ with respect to the outer corona (region III). In addition, a clear difference is  present for the two values of $c$ regarding the inner corona region (region II): while its collapse is approximately the same as that of the core for $c=5\%$ (panel e), a clear delay in $\alpha$ is present for the less crosslinked microgel (panel b). In the latter case, a clear sequence of inflection points is found when going from the inner to the outer region. Similarly to the neutral case, we still observe that the behaviour of the sphere of radius \rg roughly follows that of the core, while that of radius \rh is very close to that of the corona region. These results clearly show that the development of a minimum in the ratio between $R_g$ and $R_H$, as reported in Fig.~\ref{fig1-2}, can be easily interpreted in terms of the different collapse behaviour that involves, respectively, the core and the corona region, the latter intended as a whole. 

We thus refer to this phenomenology for simplicity as a `two-step' collapse, although the number of steps may depend on the level of heterogeneity of the underlying microgels. Its microscopic origin can be traced back to the different affinity of the charged monomers with respect to the solvent and to their different electrostatic screening conditions. Indeed, since charges remain solvophilic at all values of $\alpha$ and are accumulated in the corona, this provides the microgel with a heterogeneous collapse mechanism which differentiates the core (where charges are absent) from the corona (where most of the charges are located). 

While surface-charged microgels unambiguously show a two-step collapse for both studied $c$ values, the situation is different for randomly-charged microgels with $f=3.2\%$, reported in Figs.~\ref{fig:local_swelling}(c,f), which display a different behaviour according to the value of $c$: while for $c=1.25\%$ the outer corona is found to undergo a deswelling transition at a larger value of $\alpha$ with respect to the core, this is not the case for $c=5\%$. This is reflected in the occurrence of a minimum in \ratio only for the randomly charged case with $c=1.25\%$, while this is absent for $c=5\%$, as previously shown in Fig.~\ref{fig1-2}(f). 

To further confirm that the presence of the minimum is associated to a non-homogeneous internal deswelling of the microgels, we report the snapshots of randomly-charged microgels for the two studied $c$-values in  Fig.~\ref{fig:snaps}. Starting with the $c=1.25\%$ microgel, we clearly see that long dangling chains remain in rather extended configurations even when the internal portions of the networks have already collapsed. This is also evident at $\alpha$ values slightly above the VPT, where both core and inner corona are found in a compact spherical shape, while the outer chains are still protruding out of the sphere, giving the microgel an anisotropic global structure.
It is instructive to focus on the outer corona monomers at intermediate values of $\alpha$, where the external chains are formed by a mixture of collapsed and extended regions, due to the different solvophobic interactions of charged and neutral monomers, respectively. Although the charge distribution is random, the large amount of long chains in a low-crosslinked microgel (see Fig.~\ref{fig:chain-lengths}) makes it favourable to have several charged groups in these regions. This triggers the onset of a minimum in $R_g/R_H$, precisely by the same mechanism occurring in surface-charged microgels. However, such behaviour is not observed for $c=5\%$, where the larger amount of crosslinkers limits the chain length in the corona region, thus inhibiting the mutual influence of charges on the dangling chains, as also visible in the snapshots of Fig.~\ref{fig:snaps},  illustrating a rather homogenous deswelling for the $c=5\%$ randomly charged microgel at all values of $\alpha$. For comparison, corresponding snapshots for surface-charged microgels are reported in Fig.~\ref{fig:surfchargedmgel}.

On the basis of the above evidence, we can thus associate the occurrence of a minimum in \ratio to the two-step collapse of the microgel, differentiating core and corona behaviour. While the underlying topology is not responsible for such a mechanism, it facilitates its onset even for randomly-charged microgels at low $c$, due to the more heterogeneous nature of the network under these conditions.

\subsection*{Effect of crosslinker concentration, charge amount  and charge distribution on the minimum in \ratio}
In the previous subsection we provided evidence that the presence of charged groups is a necessary condition for the occurrence of a minimum in $R_g/R_H$, but also that this is found to depend on the specific charge distribution and on the crosslinker concentration. While for microgels decorated with surface charges for $f=3.2\%$, we observed the minimum for both studied values of $c$, which can be intuitively understood in terms of the delayed collapse of charged surface chains, the situation is more complex for randomly-charged microgels, where the presence of the minimum is found to depend also on $c$. To shed light on this aspect, we perform additional simulations of  microgels with random charge distributions and investigate the effect of varying $f$ as well as $c$.  The behaviour of \ratio as a function of effective temperature $\alpha$  is reported in Figs.~\ref{fig:random}(a-c) for a large variety of microgels, with $c$ varying between 1.25\% and 10\% and $f$ exploring a range from $1\%$ to $10\%$.

We observe that, upon decreasing $c$ while leaving $f$ unchanged, the minimum becomes more and more evident in agreement with unpublished experimental results~\cite{sun2005investigation}. In addition, Fig.~\ref{fig:random} shows that the minimum is more pronounced with increasing $f$ at the same value of $c$, also shifting to larger and larger values of $\alpha$. In all cases, the decrease of \ratio at the minimum is also accompanied by its decrease at small $\alpha$, indicating that the growth of $R_H$ is not compensated by that of $R_g$ under swollen conditions. At large $\alpha$ most curves tend to recover the HS value, except for the most charged microgels, that are expected to completely collapse at  much larger effective temperatures~\cite{capriles2008coupled,del2020charge}, beyond those explored in the present simulations.

\begin{figure}[]
	\centering
	\includegraphics[width=0.5\textwidth]{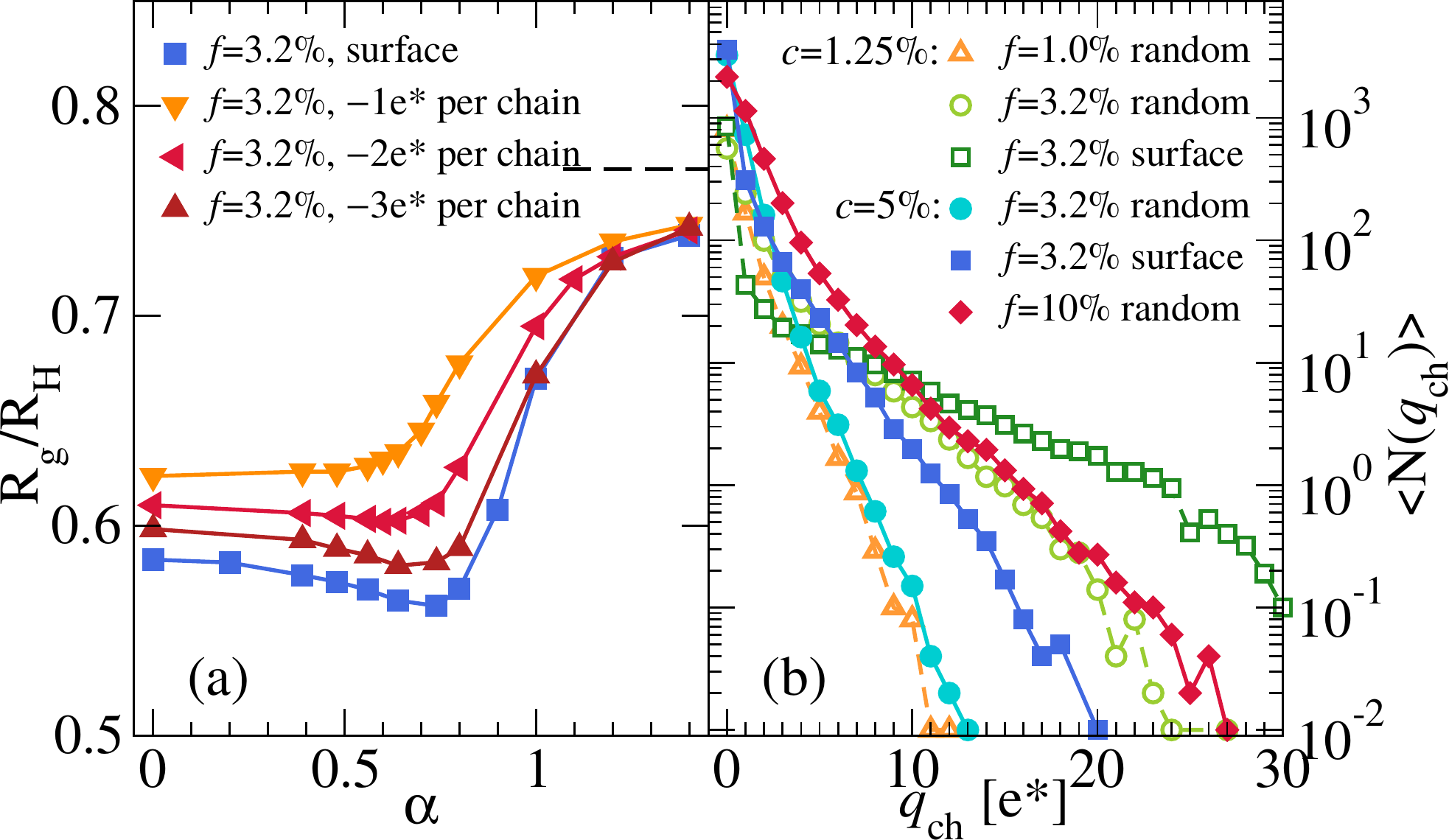}
	\caption{\label{fig:surface} 
		\textbf{Ratio \ratio for c=5\% surface-charged microgels (a) and distribution $N(q_\text{ch})$ of number of charges per chain (b) } Surface-charged microgels are shown with filled symbols, while open triangles indicate microgels with $f=3.2\%$ for which charges are constrained to be found as one, two or three charges on each external chain, indicated as $-1e^*, -2e^*, -3e^*$ per chain, respectively. }
\end{figure}

From all data in Fig.~\ref{fig:random}(a), we can build a plot reporting the depth of the minimum $\Delta_\text{min}$ (defined in Methods) as a function of charge content, reported in Fig.~\ref{fig:random}(d), for different values of $c$, identifying regions in the $(c,f)$ parameter space where no minimum is observed (homogeneous deswelling), as opposed to others where this is present (two-step deswelling). Clearly, no minimum is found for low $c$ and low $f$. However, at very large $c$, the minimum is also inhibited for very large charge content.
Finally, the results in Fig.~\ref{fig:random} allow us to exclude the presence of a minimum in \ratio for highly cross-linked charged microgels. 

Having shown that the minimum arises also in randomly-charged microgels above a certain charge fraction for each considered $c$, we now further clarify the role of the specific charge distribution by tuning it \textit{ad hoc}.
To this aim, we investigate $c=5\%$ microgels where charges are located on the surface but with the additional constraint that a fixed number of charges per (surface) chain should be present, varying this number from one to three. In this way we want to assess the role played by charge correlations and whether there exists a minimum amount of charges per chain required to observe the minimum in $R_g/R_H$. The behaviour of \ratio for these cases is reported in Fig.~\ref{fig:surface}(a), where we clearly find confirmation that there is no minimum when only one charge per chain is present, even if they are all located on the surface (as shown by the charged monomers radial distributions in Fig.~\ref{fig:ions-profs-surface}). A tiny minimum appears when two charges per surface chain are present and finally a well-defined minimum occurs for three charges per chain. The systematic increase of the number of charges per surface chain makes the behaviour of \ratio more and more similar to the originally considered case of surface charges that are randomly distributed throughout the corona, thus mimicking the experimental situation of a pNIPAM microgel with a given amount of initiator arranged on the outer part of the microgel surface in a disordered fashion. Not only the minimum becomes more pronounced, but its position also moves towards larger effective temperatures. 
Additional data for $f=10\%$ and ``mixed'' charge distribution (partially located on the surface and partially random) are reported in the SI (Fig.~\ref{fig:rg-rh-surface}).

\begin{figure*}[]
\centering
	\includegraphics[width=0.85\textwidth]{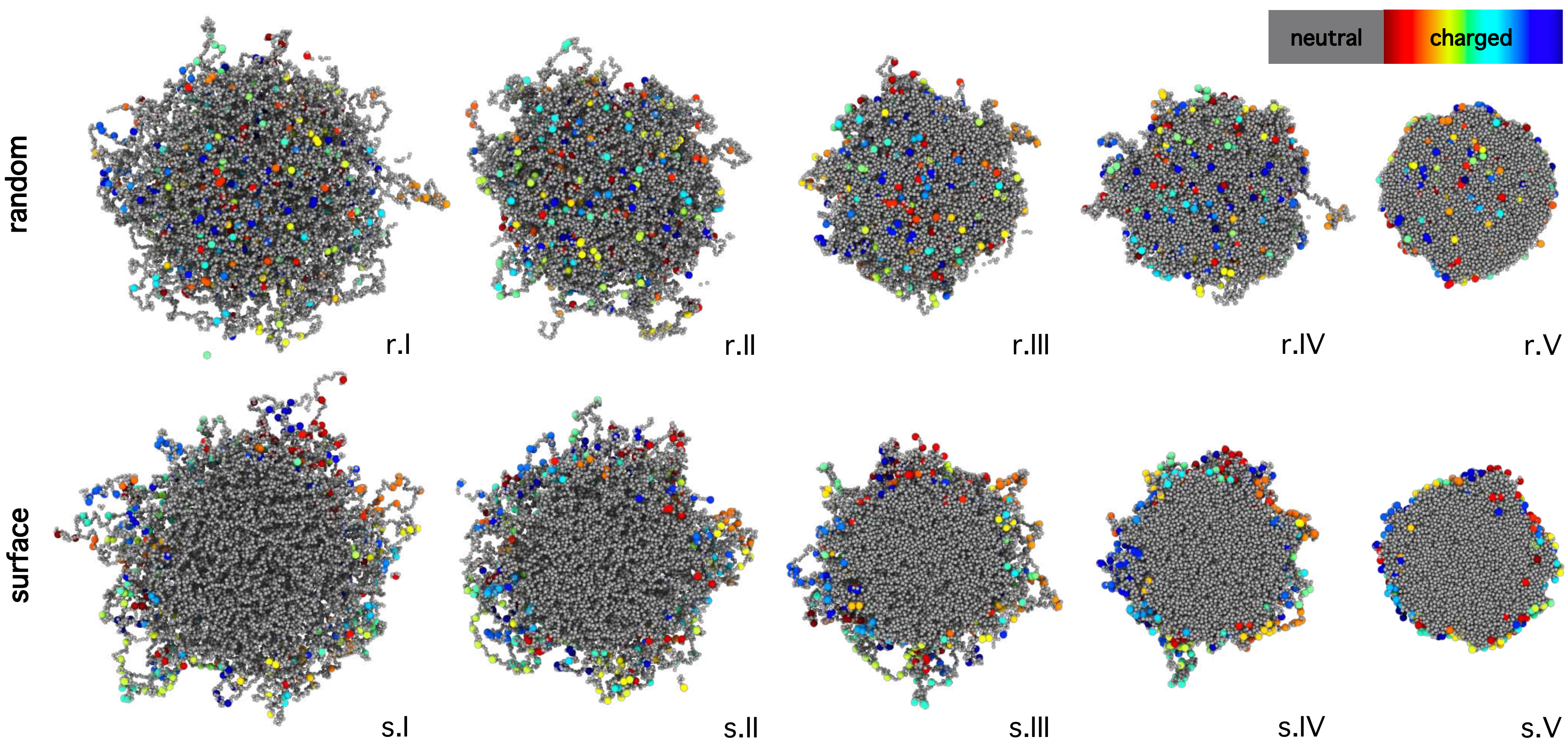}
	\caption{\label{fig:charges_snaps} 
		\textbf{Snapshots highlighting charged beads.} 
		Simulation snapshots representing cross-sections of microgels with $f=3.2\%$ and $c=5\%$ with random (top row) and surface (bottom row) charge distribution referring to different values of the swelling parameter
		$\alpha=0, 0.48, 0.74, 0.80, 1.20$ from the swollen to the collapsed state (from left to right).
		Charges belonging to the same chain are shown with the same color and enhanced size, to improve their visualisation, while uncharged monomers are grey. }
\end{figure*}

It is important to notice that, in the case of standard surface charge distribution with $f=3.2\%$, the average charge per surface chain is roughly equal to $\sim -1.4e^*$ only. However, the minimum is much more pronounced in this case than when the charge is set to be  $-3e^*$. In this case, the variance of the number of charges per chain is not negligible, as displayed in Fig.~\ref{fig:surface}(b). In this plot, we report the distribution of the number of charges per chain for random and surface charge distributions with varying $c$ and $f$, showing that all curves referring to microgels not displaying the two-step collapse are rather well-described by a single exponential decay, while those with a well-developed minimum reveal the onset of a tail in the distribution. This suggests the development of a characteristic correlation between charges on long chains, which makes the heterogeneous deswelling mechanism much more efficient. The standard deviation of the distribution of charges per chain is further enhanced at $c=1.25\%$, due to the underlying network topology, making the minimum in \ratio always more pronounced for low-crosslinked microgels than for the corresponding ones at higher $c$.

To visualize what happens at the level of single chains, we show in Fig.~\ref{fig:charges_snaps} the snapshots of microgel cross-sections for different values of $\alpha$. In particular, we compare random and surface charge distributions with $f=3.2\%$ and $c=5.0\%$, and we draw with the same color charges belonging to the same chain. In this way, we aim to visualize the presence of multiple charges per chain and how the deswelling process is affected. We see that for randomly-charged microgels the charged chains, despite a few isolated cases,  play a minor role in the overall collapse of the network, probably because they are rare and pulled by the neighboring neutral monomers towards the collapsed network with increasing $\alpha$. Instead, for surface charged microgels, there are many more charged chains, each in turn with several charged monomers which prevent the chains from a full collapse. These microgels are thus found to maintain an extended conformation up to large values of $\alpha$, even beyond the VPT. Hence, prior to the final collapse at very large $\alpha$, the microgel clearly looks like an inner compact sphere decorated by many non-compact chains (see, for instance, Fig.~\ref{fig:charges_snaps}(s.IV)).

\subsection*{Predicting the two-step collapse: a simple indicator}
\begin{figure}[]
	\centering
	\includegraphics[width=0.5\textwidth]{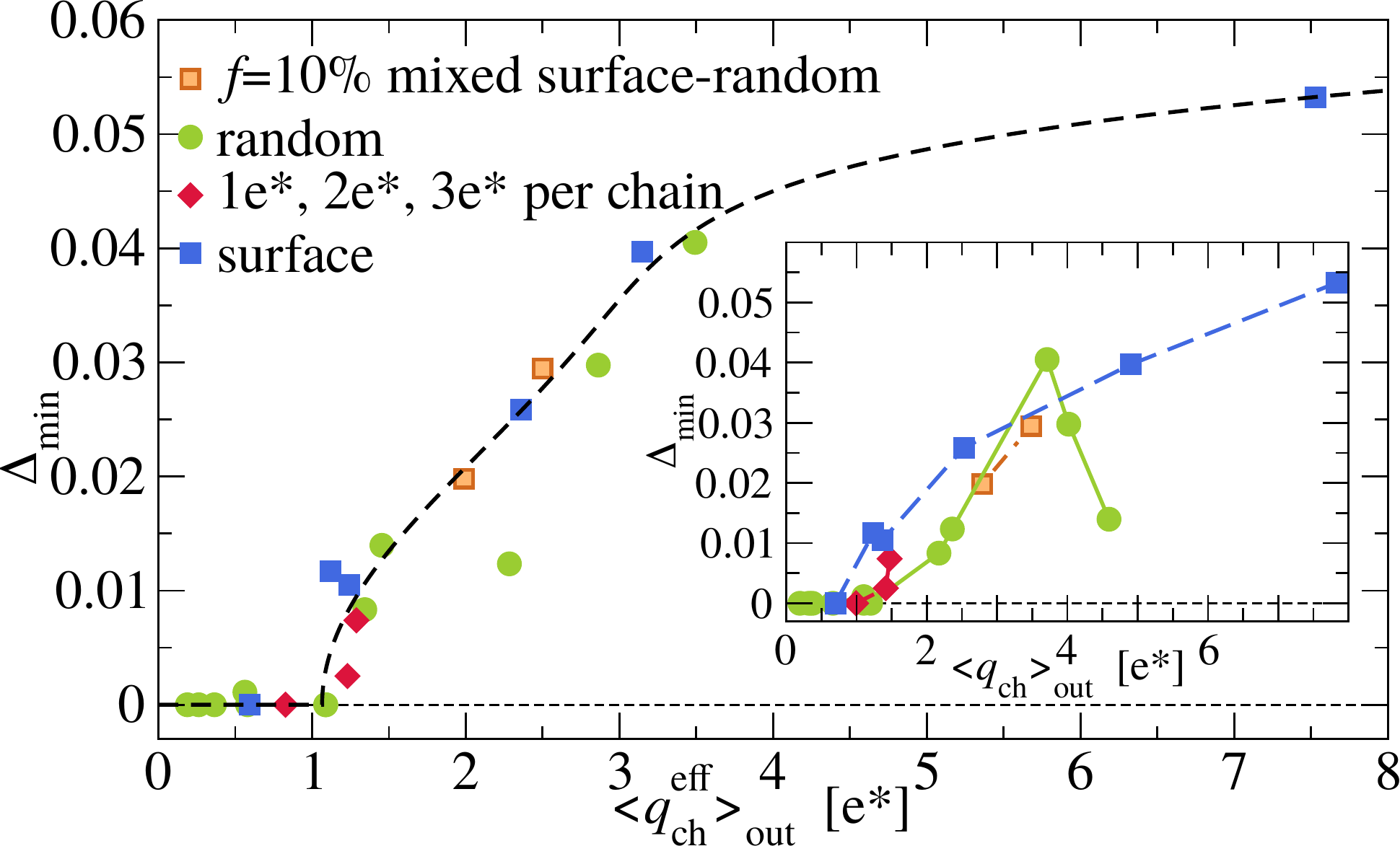}
	\caption{\label{fig:qout} 
		\textbf{Prediction of two-step collapse.} Master plot for the minimum $\Delta_\text{min}$ as a function of the average effective charge $\qchEffOut$ of the surface chains. 		Lines are guides to the eye. Inset: same for the bare charge $\qchOut$.	}
\end{figure}
To take full advantage of the predictive ability of the simulations, we are now in the position to establish an indicator that may be useful in predicting the appearance of the minimum in $R_g/R_H$. Since we demonstrated that the two-step collapse is due to the explicit presence of charged groups in the microgels and preferentially to their arrangement on the microgel corona, we propose to identify the average number of charges per chain on the microgel surface, defined as $\qchOut$ as the observable that controls the onset and the development of the minimum. To verify that this is the case, we plot  the dependence of the minimum depth $\Delta_\text{min}$ as a function of $\qchOut$ in the inset of Fig.~\ref{fig:qout}. We find that microgels with the same charge distribution, either surface or random, independently of their values of $c$ and $f$, display characteristic behaviours that are roughly monotonic in $\qchOut$, except when this is very large in the case of randomly charged microgels (see below).  We also notice that no minimum is observed when $\qchOut \lesssim1$, meaning that on average more than one charge per surface chain is needed in order to develop the onset of the two-step collapse.
Then, as $\qchOut$ increases, the depth of the minimum gets more and more pronounced both for surface-charged microgels and for those where we impose a fixed number of charges per chain. A similar behaviour is observed for microgels with random charges at low $f$ content, and thus low $\qchOut$. However, when this increases, we find first a sudden growth of $\Delta_\text{min}$, which even exceeds that of surface-charged microgels at the same $\qchOut$, followed by a drastic reduction and a highly non-monotonic behaviour, calling for a further refining of the chosen indicator.

To this aim, we exploit the explicit charge treatment in our simulations and consider the screening effect of the counterions, so that rather than the bare charge, we consider an effective charge per surface chain $\qchEffOut$ (see Methods and SI).
Therefore, in Fig.~\ref{fig:qout} we plot the minimum depth $\Delta_\text{min}$ as a function of $\qchEffOut$. Remarkably, a master plot is found  where data for all simulated microgels ($24$ different systems in total) fall on the same curve manifesting the same behaviour. We notice that a couple of points are found to slightly deviate from the master curve, which may  well be attributed to the statistical uncertainty of our calculations, which are carried out for a single topology and charge realization for each microgel.
However, it is noteworthy that the use of the effective charge is able to completely remove the non-monotonicity of the data for randomly charged microgels as a function of bare charge, indicating the overall correctness of our procedure. In this respect, we remark that this is due to the fact that screening effects are much more effective for random rather than for surface charged microgels, which can be intuitively understood in terms of an overall reduced electrostatic repulsion experienced by the counterions when charges are distributed throughout the entire network, thus facilitating the microgel collapse.

\section*{Discussion}

In this work, we thoroughly investigated the swelling behaviour of microgels across the VPT upon varying crosslinker concentration and charge distribution. The study was motivated by experimental measurements of gyration and hydrodynamic radii that showed the appearance of a characteristic minimum in the ratio \ratio at a temperature close to the VPT one.
Such a minimum indicates that a different rate of collapse with increasing temperature takes place between the inner mass of the microgel, quantified in size by the gyration radius, and its outer mass, associated to the hydrodynamic radius. A similar evidence was also reported in few previous works~\cite{arleth2005volume,sun2005investigation}, but its microscopic origin was not investigated in detail so far. The new measurements reported in this paper for four different samples unambiguously reveal the presence of such a minimum.
In a previous experimental work by some of us~\cite{truzzolillo2018overcharging}, the minimum was tentatively associated to the presence of initiator charges, manifesting their effect under collapsed conditions even for simple pNIPAM microgels. This conjecture was supported by electrophoretic mobility measurements also displaying a rapid increase of the signal for temperatures above the VPT, thus providing evidence that, in the high temperature regime, pNIPAM microgels have to be considered as charged micro-objects~\cite{scotti2016role,braibanti2016impact,sennato2021double}.
However, a legitimate doubt concerned whether the minimum in \ratio would be solely due to charges or could be attributed to other ingredients, such as the inhomogeneous core-corona topology of pNIPAM microgels synthesized by precipitation polymerisation methods. For these reasons, the primary purpose of the present work was to correctly assess the role of charges on the occurrence of the minimum. 

We thus performed an extensive set of simulations for several realistic microgels, with different crosslinker concentration, charge content and distribution. First of all, we found that the minimum does not occur for purely neutral microgels, even for very low cross-linked concentrations (like $c=1.25\%$, the lowest value investigated in this work), where the inhomogeneous internal structure of the microgels is enhanced.
Hence, we can safely exclude that the two-step collapse is associated to the topology, whereas it must be attributed to additional interactions emerging at the VPT which modify the homogeneous deswelling that would take  place in a perfectly neutral microgel.  Incidentally, it is important to notice that, to our knowledge, it is currently not possible to synthesize strictly uncharged microgels made of pNIPAM, the only example being for microgels made of (non-thermoresponsive) polystyrene~\cite{van2017fragility}, thus preventing confirmation of the present numerical predictions. Alternatively, experiments could be performed in the presence of added salt that would screen electrostatic interactions~\cite{braibanti2016impact}, though at the same time favoring microgel-microgel aggregation at high temperatures.

The additional mechanism giving rise to the two-step deswelling is thus attributed to electrostatic effects. In particular, charged initiator groups  interact with the solvent in a different way with respect to NIPAM monomers as temperature increases, since they always maintain a solvophilic character, acting against collapse. When the amount of initiator is very low as in common synthesis procedures, these effects can be very subtle and may depend on the location of these charges. Therefore, it is important to discriminate between random location of charges throughout the network and surface arrangement at the periphery of the corona.
Indeed, the latter case seems to be appropriate to model charged initiators which start the polymerisation process and then possibly remain confined in the outer part of the microgel to minimize mutual electrostatic interactions. In this way, being localized on the surface, upon particle collapse the charged groups would experience a strong mutual repulsion, thus influencing the corona deswelling as demonstrated in the previous sections. On the other hand, a random distribution of charges would `dilute' the effect of the electrostatic contribution throughout the whole particle, including the core, unlikely affecting the collapse of the particle during the VPT.

The simulation results presented here clearly show that, for typical amounts of initiators used in synthesis of pNIPAM microgels ($f=3.2\%$) and standard crosslinker concentration (c=5\%), a random distribution of charges does not give rise to a minimum in $R_g/R_H$, while this is present when charges are distributed only on the surface. Actually, the minimum for random charge arrangement only manifests at much higher $f$ or lower $c$, a numerical prediction that awaits for experimental confirmation. In this respect, it would be useful to perform measurements of $R_g/R_H$ for ionic microgels, e.g. in the presence of a second component such as PAAc. In this case, due to different reaction rates, the co-monomers might still be inhomogeneously distributed~\cite{stefanie}, so it remains to be verified whether a random distribution of charges may still apply in certain cases. 

On the other hand, our simulations showed that microgels with surface charge distribution seem to be the appropriate model to describe the occurrence of the minimum in pNIPAM microgels, in agreement with experimental measurements and with previous hypothesis~\cite{howe,zhou2012correlation,daly2000temperature}, although a direct proof of charge location in experiments would be very valuable.  To this purpose, super-resolution microscopy experiments~\cite{conley2017jamming} could be helpful when combined with selective labeling of the surface or even of the initiators, as well as studies where  the average bare charge and screening electrostatic conditions of the polymer chains would be systematically varied, as recently done in Ref.~\cite{bergman2020controlling}.
Additionally, one could rely on neutron scattering measurements, where differentiations of the molecular components could be achieved by contrast variation~\cite{mohanty2017interpenetration,nojd2018deswelling,scotti2018hollow}.
The numerical results of the present work will further provide a framework to interpret experimental data thanks to the master curve reported in Fig.~\ref{fig:qout}, which makes use of a single quantity, the screened average charge per surface chain $\qchEffOut$, to describe the two-step deswelling process. These experiments could also be used to estimate such an effective charge.

Another important point to be addressed in the future is to systematically vary the amount of initiator, a relatively unexplored  parameter. It will be interesting to see to which extent it can be exploited, given that too large amount of initiator could give rise to synthesis by-products that would not necessarily be part of the microgel network. In addition, it could be possible that above some threshold, the charges would start to penetrate in the interior of the microgel giving rise to `mixed' surface-random distribution, as discussed in Ref.~\cite{huang2017hollow}. Indeed, looking at the experimental  behavior  for $I_\text{KPS}$=8\% (Fig. \ref{fig1-2}),  we observe  that this sample displays larger values of $R_g/R_H$ at low temperatures and a shallower minimum at the VPT, in qualitative agreement with the effect of mixed charge distributions shown in the SI. These findings seem to suggest that increasing the initiator molar fraction in the synthesis does not simply give rise to an augmented charge distribution peaked at the microgel periphery, but presumably homogenizes the initiator (and possibly the mass) distribution radially within the microgel. 

Finally, the outcomes of this work will also be relevant for the study of the collective behaviour of pNIPAM microgels at high temperatures, where numerical/theoretical works are currently lacking. While it is now established that the effective potential for swollen microgels can be described as a soft repulsion~\cite{rovigatti2019elasticity,bergman2018new}, at high temperatures attractive interactions should become important~\cite{howe,schweizer2021viscoelasticity}, albeit mediated by the electrostatic interactions of the initiator. It will be therefore necessary to appropriately assess their contribution to the collective and rheological behaviour of dense microgel suspensions.

\bigskip
\small
\section*{Models, Materials and Methods}
\subsection*{Synthesis} 
Surfactant-free radical polymerisation has been used to synthesize different batches of pNIPAM microgels. Specifically, we used methylene-bis-acrylamide (BIS) as crosslinker agent and two cationic initiators: potassium persulfate (\ch{K2S2O8}, KPS) and ammonium persulfate (\ch{(NH4)2S2O8}, APS).
Three different batches with $c = 5.25$\si{\% \mole} of BIS  have been synthesized with the following initiator content and type: 
$1.6$\si {\% \mole} of KPS;  $1.6$\si {\% \mole} of APS; $8.0$\si {\% \mole} of KPS.
Since our main focus is to scrutinize the role of initiator charges, we primarily use microgels obtained via surfactant-free emulsion polymerization, which allows us to minimize the impact of residual ionic surfactants on microgel deswelling. However, we also carried out one synthesis in the presence of surfactant and used the sample after purification, in order to test the possible impact of surfactant addition, while at the same time reducing both the microgel size for low crosslinker content and the associated error on the $R_g$ measurement.  In this way, we obtained a batch of smaller particles, which has been prepared through the addition of an amount of $0.85$\si{\% \mole} of sodium dodecyl-sulfate (SDS). In this case microgels are synthesized with $c = 1.37$\si{\% \mole} of BIS and $1.6$\si{\% \mole} of KPS.
All samples have been further purified through 3 consecutive cycles of centrifugation and supernatant replacement with deionized water. Finally, 2 \si{\milli\mole} of sodium azide \ch{NaN3} have been added to the final samples to prevent bacterial growth.

\subsection*{\label{subsec:experiments}Experiments} 
To characterize the structure of particles in suspension we employed static and dynamic light scattering (SLS, DLS).
For all the samples an aliquot of the mother batches ($1$\si{wt\percent} of NIPAM) has been diluted 60 times so to reach a mass fraction of $0.017$\si{wt\percent} of NIPAM, equivalent to microgel volume fractions lower than 0.4\% as determined by viscosimetry following the method detailed in Ref.~\cite{truzzolillo2015bulk}. All scattering experiments have been performed by using the same laser source ($\lambda=532$\si{\nano\meter}). Before each data acquisition, each sample has been thermostatted for $20$\si{\minute} via a recirculating bath with $0.1$\si{\celsius} accuracy and light has been collected by varying the scattering angle for SLS measurements with \ang{1} accuracy.
We extracted the average gyration radius of microgels, fitting the low scattering-vector part of the intensity $I(q)$ with the functional form (Guinier approximation) as discussed in the SI (see Fig.~\ref{fig:Ik-gtk}). For all samples data could be reliably fitted in the entire range $5.5$\si{\micro\meter^{-1}}$\le q \le 2/R_g$. The uncertainty on Rg is given by the fit error, the latter being less than $10\%$ of the best-fit value.
For all the samples, $I(q)$ has been measured over a range of scattering wave-vectors going from $q = 5.55$\si{\micro\metre^{-1}} to $q = 30 $\si{\micro\metre^{-1}}.

Through dynamic light scattering we measured the autocorrelation function of the scattered intensity at a fixed value of $q=q_\text{DLS}$. At high dilution, this can be put in relation with the square of the self intermediate scattering function ${F_s(\vec{q},t)}^2$, which for simply diffusive systems has the functional shape of a convolution of exponential decays, taking into account the effects of polydispersity.
The analysis of ${F_s(\vec{q},t)}^2$ has been performed by cumulant analysis, up to the second central moment of the distribution of decay rates:
\begin{equation}\label{eq:cumulants}
{F_s(\vec{q}_\text{DLS},t)}^2 \propto e^{-2q_\text{DLS}^2\bar{D}t} \left( 1 + \frac{\mu_2 t^2}{2!} + o(t^3) \right)^2
\end{equation}
where $\bar{D}$ is the average self-diffusion coefficient and $\mu_2$ is related to the second moment of the distribution of $D$ of the molecules in the sample.
Using Stokes formula we obtain the average hydrodynamic radius as $R_H=k_B T/(6\pi\eta_0 \bar{D})$, and the corresponding dispersion is $\sigma_{R_H}=\sqrt{\mu_2}R_H/(\bar{D}q_\text{DLS}^2)$, where $k_B$ is the Boltzmann's constant, $T$ the bath temperature and $\eta_0$ is the zero-shear viscosity of the solvent.
For microgels with $c=1.37\%$ we measured the intensity time-correlation function at $q_\text{DLS} = 22\si{\micro\metre^{-1}}$, corresponding to an angle of \ang{90}, while for microgels with $c=5.25\%$ we gathered data at both $q_\text{DLS} = 22\si{\micro\metre^{-1}}$ and $q_\text{DLS} = 15.7\si{\micro\metre^{-1}}$ (the latter corresponding to an angle of \ang{60}) in order to possibly detect and reject spurious effects of polydispersity when $q_\text{DLS}$ was close to a value correspondent to the first minimum of the microgel form factor. The polydispersity indexes of the samples $\text{PDI} = \mu_2/(\bar{D}^2q_\text{DLS}^4)$, calculated at $T\simeq25$\si{\celsius}, are: (i) $0.065$ for $c = 1.37$\si{\% \mole} and $I_\text{KPS} = 1.6$\si{\% \mole}, (ii) $0.148$ for $c = 5.25$\si{\% \mole} and $I_\text{KPS} = 8.0$\si{\% \mole}, (iii) $0.074$ for $c = 5.25$\si{\% \mole} and $I_\text{KPS} = 1.6$\si{\% \mole}, and (iv) $0.005$ for $c = 5.25$\si{\% \mole} and $I_\text{APS} = 1.6$\si{\% \mole}.

\subsection*{\label{subsec:simulations}Numerical simulations}
We simulate  monomer-resolved microgels of $N\sim42000$ beads exploiting a recently developed protocol~\cite{gnan2017in} able to generate fully-bonded, disordered polymer networks. By appropriately tuning the internal polymer distribution, the model faithfully reproduces measured swelling properties and  form factors of neutral microgels across the VPT~\cite{ninarello2019modeling}. The resulting network structure is made of equal-size beads, representing polymer segments interacting with the well-established Kremer-Grest potential~\cite{grest1986molecular}. In this model, all beads experience a steric repulsion, modeled with the Weeks-Chandler-Anderson (WCA) potential:
\begin{equation}
\label{eq:wca}
V_{\text{WCA}}(r)  =  
\begin{cases}
4\epsilon\left[\left(\frac{\sigma}{r}\right)^{12}-\left(\frac{\sigma}{r}\right)^6\right]+\epsilon & \quad \text{if} \quad r \le 2^{1/6}\sigma  \\
0 & \quad \text{if}  \quad  r > 2^{1/6}\sigma
\end{cases}
\end{equation}
where $\epsilon$ and $\sigma$ are the energy and length units, respectively. Connected beads also interact via  the finitely extensible nonlinear elastic potential (FENE):
\begin{equation}
V_{\text{FENE}}(r)  = 
-\epsilon k_F{R_0}^2\log\left[1-{\left(\frac{r}{R_0\sigma}\right)}^2\right], \quad r < R_0\sigma 
\end{equation}
where $R_0=1.5\sigma$ is the maximum bond distance and $k_F=15$ is a stiffness parameter influencing the rigidity of the bond. These bonds cannot break during the simulation, mimicking strong covalent bonding. 
Beads that are linked via FENE to two neighbours represent NIPAM monomers, while those mimicking crosslinkers  have fourfold valence. We analyse in detail values of the crosslinker fraction similar to the experimental ones, i.e. $c = $1.25\% and 5.0\%. In order to be more quantitative on the $c$-dependence of results, we also provide specific additional results for  $c=$ 3.0\%, 7.5\% and 10\%. 

We consider that microgels obtained via free-radical polymerisation embed a certain amount of ionic groups in their backbone, in this case \ch{SO4-}, resulting from the dissociation of APS/KPS after a further thermal splitting of the ions \ch{S2O8^2-}, starting the reaction and remaining attached to the network. This is modeled by providing a fraction $f$ of the beads with a negative charge and inserting an equivalent number of counterions with positive charge to preserve overall electro-neutrality. Counterions sterically interact among each other and with microgel beads through the WCA potential. Their diameter is set to $\sigma_c=0.1\sigma$ to avoid spurious effects from excluded volume~\cite{del2019numerical}, while electrostatic interactions are given by the Coulomb potential $V_{\rm coul}(r_{ij}) = \frac{q_i q_j \sigma}{e^{*2} r_{ij} } \epsilon$, where $q_i$ and $q_j$ are the charges of the beads, and $e^* = \sqrt{4\pi\varepsilon_0\varepsilon_r\sigma\epsilon}$ is the reduced unit for the charge, embedding the vacuum and relative dielectric constants, $\varepsilon_0$ and $\varepsilon_r$. $q_i$ and $q_j$ are set to the values $-e^*$ or $+e^*$ whether it refers to charged beads or counterions.
For different values of $c$ we consider several values for the fraction of charged monomers.  We analyse in detail the case $f=3.2\%$,  corresponding to an initiator concentration $I_\text{KPS} = 1.6\%\si{mol}$, that matches the nominal charge content in some of the performed experiments. We also consider $f=10\%$, a relatively high charge fraction that should mimic the experimental situation with a nominal initiator content of $I_X = 8.0\%\si{mol}$, assuming that in a synthesis process with a large quantity of highly reactive initiator radicals, a non-negligible amount of charged groups would end up in low molecular mass byproducts, that will be removed by filtration at the end of the process. For comparison, we also study the neutral case ($f=0$) to highlight effects entirely attributable to charges. Simulations are performed for both random and surface charge distributions. In the former case, charged beads are randomly distributed throughout the network (except on crosslinkers), while for the latter they are found only on surface chains. To this aim, we randomly choose $f$ beads located in the exterior corona, whose distance from the microgel centre of mass is greater than $R_g$.
In all simulations the solvent is implicitly taken into account through an effective potential which mimics the change in the polymer-solvent affinity by raising temperature:
\begin{equation}\label{eq:valpha}
V_{\alpha}(r)  =  
\begin{cases}
-\epsilon\alpha & \text{if } r \le 2^{1/6}\sigma  \\
\frac{1}{2}\alpha\epsilon\left\{\cos\left[\gamma{\left(\frac{r}{\sigma}\right)}^2+\beta\right]-1\right\} & \text{if } 2^{1/6}\sigma < r \le R_0\sigma  \\
0 & \text{if } r > R_0\sigma
\end{cases}
\end{equation}
where $\alpha$ is the solvophobicity parameter representing the effective temperature, which is varied from $\alpha=0$ (good solvent conditions) to $\alpha\gtrsim 1.40$ (bad solvent conditions) to reproduce the swelling curve from the swollen state to the most collapsed ones. Here $\gamma = \pi\left(\frac{9}{4}-2^{1/3}\right)^{-1}$ and $\beta = 2\pi - \frac{9}{4}\gamma$ are constants defining the functional shape of the potential~\cite{soddemann2001generic}. We showed in a previous work~\cite{del2020charge} that, for a correct description of charge effects, only neutral monomers should interact with the additional $V_{\alpha}$ potential. Instead, charged monomers retain a solvophilic character at all temperatures, i.e. for the interaction among charged beads and other charged or neutral ones we always set $\alpha=0$.
The equations of motion are integrated through a Nos\'e-Hoover thermostat in the constant NVT ensemble for the equilibration, and through a Velocity-Verlet algorithm in the constant-energy ensemble for the production runs, with an integration time-step $\Delta t = 0.002\tau$, where $\tau = \sqrt{m\sigma^2/\epsilon}$ is the reduced time unit.  All simulations are performed with the LAMMPS package~\cite{LAMMPS} at fixed temperature $k_BT/\epsilon = 1.0$.
Long-range Coulomb interactions are computed with the particle-particle-particle-mesh method \cite{p3m}. The equilibration of each system is carried out for $2000\tau$, followed by a production run of $6000\tau$ from which we extracted the equilibrium averages of the main observables of interest, such as the gyration radius \rg and the density profile, defined as the average density at a fixed distance from the centre of mass, $\rho(r)= \left\langle \frac{\sum_{i_{=1}}^{N_s} \delta (|\vec{r}_{i}-\vec{r}_\text{cm}|-r)}{N_s} \right\rangle$.

To estimate $R_H$ in simulations, we instantaneously approximate the microgel as an effective ellipsoid~\cite{rovigatti2019elasticity} and then calculate the hydrodynamic friction using the approach developed by Hubbard and Douglas~\cite{hubbard1993hydrodynamic}. For each configuration, we first compute the convex hull containing all the beads of the microgel, then evaluate the gyration tensor of all the simplices it is made of and finally determine the ellipsoid with the same gyration tensor~\cite{rovigatti2019elasticity}.
The hydrodynamic friction $\zeta$ of each ellipsoid is calculated as~\cite{hubbard1993hydrodynamic}:
\begin{equation}
\zeta = 6\pi\eta C_\Omega \equiv 6\pi\eta R_H
\label{eq:friction}
\end{equation}
where $\eta$ is the solvent viscosity and $C_\Omega$ is the electrostatic capacitance. 
Eqn.~\ref{eq:friction} is based on a solution of the Navier-Stokes equation for steady flow of rigid particles with stick boundary conditions, where the hydrodynamic interactions are described by the isotropic angular averaged Oseen tensor~\cite{hubbard1993hydrodynamic}. In this approximation, the Navier-Stokes equation for the momentum flux density taked the same form of the Poisson's equation for electrostatics, thereby the hydrodynamic radius becomes mathematically equivalent to the electrostatic capacitance, which for rigid ellipsoids is~\cite{hubbard1993hydrodynamic}:
\begin{equation}
R_H=C_\Omega = 2\left[ \int_{0}^{\infty}\frac{1}{\sqrt{(a^2+\theta)(b^2+\theta)(c^2+\theta)}}d\theta \right]^{-1}
\label{eq:capacitance}
\end{equation}
where $a$, $b$, $c$ are the principal semiaxes.  More details on the calculation of $R_H$ are provided in the SI. 

To evidence the differences in the local structure of the microgels, we calculated the local swelling curve as the volume of three different regions, defined on the basis of the quantity of monomers they contain. Ordering the polymer beads according to their distance from the centre of mass, at each time we associate the most inner fraction of monomers $f_\text{m,I} = 0.65$ of beads to the core (region I), the subsequent fraction $f_\text{m,II} = 0.20$ of them to the inner corona shell (region II), and the farthest fraction of $f_\text{m,III} = 0.15$ beads to the outer corona shell (region III). We choose the fraction $f_\text{m,I}$ in such a way that the core region has an overall constant density profile at all $c$ and $\alpha$ values. The values of $f_\text{m,II}$ and $f_\text{m,III}$ are chosen in such a way that the monomers density profile at the boundary separating the two corona shells takes a value close to half the average density of the core region, to separate the more crosslinked inner part of the corona from the most external region with low density dangling chains (see Fig.~\ref{fig:local-profiles}). The volumes of the three regions, and their outer surfaces, are calculated as the convex hulls enclosing all the beads associated to them.

To quantify the distributions of charges per chain, we calculate the chain length distribution, where each chain is defined as the sequence of monomers between crosslinkers, and we count the number of charged monomers for each chain ($q_\text{ch}$).  To build the master plot in Fig.~\ref{fig:qout}, we define the extent of the minimum $\Delta_\text{min}$ as the difference between the value of \ratio at $\alpha=0.5$ and at its minimum.  We also define the average number of charges per chain on the microgel surface $\qchOut$, where $q_\text{ch}$ is averaged only over chains whose distance from the microgel's centre of mass is greater than $R_g$. To take into account screening effects, we also consider the effective charge per surface chain $\qchEffOut$, that is obtained by considering an average screened charge based on the integration of the ion-counterion pair-distribution function up to a determined distance, as discussed in the SI (see Fig.~\ref{fig:gr-ions-cions}).

\normalsize

\begin{acknowledgments}
	We thank E. Buratti, E. Lattuada, J. Ruiz-Franco, P. Schurtenberger and F. Sciortino for valuable discussions. We acknowledge financial support from the European Research Council (ERC Consolidator Grant 681597, MIMIC), from the European Union's Horizon 2020 research and innovation programme (Grant 731019, EUSMI), from MIUR (FARE project R16XLE2X3L, SOFTART), and from the Agence Nationale de la Recherche (ANR) (Grant N$^\circ$ ANR-20-CE06-0030-01, THELECTRA). The authors gratefully acknowledge the computing time granted by EUSMI on the supercomputer JURECA at the J\"ulich Supercomputing Centre (JSC).
\end{acknowledgments}

\section*{Competing interests}
The authors declare no conflict of interest.\\

\clearpage

%%%%%%%%%% Merge with supplemental materials %%%%%%%%%%
\pagebreak
\widetext
\begin{center}
	\textbf{\LARGE Supplemental Materials: Two-step deswelling in the Volume Phase Transition of thermoresponsive microgels}\\
	\bigskip
	{Giovanni Del Monte,\textsuperscript{1,2,3} Domenico Truzzolillo,\textsuperscript{4,*} Fabrizio Camerin,\textsuperscript{2,1} Andrea Ninarello,\textsuperscript{2,1} Edouard Chauveau,\textsuperscript{4} Letizia Tavagnacco,\textsuperscript{2,1} Nicoletta Gnan,\textsuperscript{2,1} Lorenzo Rovigatti,\textsuperscript{1,2} Simona Sennato,\textsuperscript{2,1} Emanuela Zaccarelli\textsuperscript{2,1,*}}\\
	\bigskip
	\footnotesize{Corresponding authors (*): domenico.truzzolillo@umontpellier.fr, emanuela.zaccarelli@cnr.it}
%	\correspondingauthor{Corresponding Author Name.\\E-mail: domenico.truzzolillo@umontpellier.fr, emanuela.zaccarelli@cnr.it}
	
\end{center}
%%%%%%%%%% Merge with supplemental materials %%%%%%%%%%
%%%%%%%%%% Prefix a "S" to all equations, figures, tables and reset the counter %%%%%%%%%%
\setcounter{equation}{0}
\setcounter{figure}{0}
\setcounter{table}{0}
\setcounter{page}{1}
\makeatletter
\renewcommand{\theequation}{S\arabic{equation}}
\renewcommand{\thefigure}{S\arabic{figure}}
\renewcommand{\thetable}{S\arabic{table}}
%\setcounter{section}{0}
%\renewcommand{\thesection}{S-\Roman{section}}
%\renewcommand{\bibnumfmt}[1]{[S#1]}
%\renewcommand{\citenumfont}[1]{S#1}
%%%%%%%%%% Prefix a "S" to all equations, figures, tables and reset the counter %%%%%%%%%%

\section*{Swelling ratio and transition temperatures}
\begin{figure}[h]
	\centering
	\includegraphics[width=\textwidth]{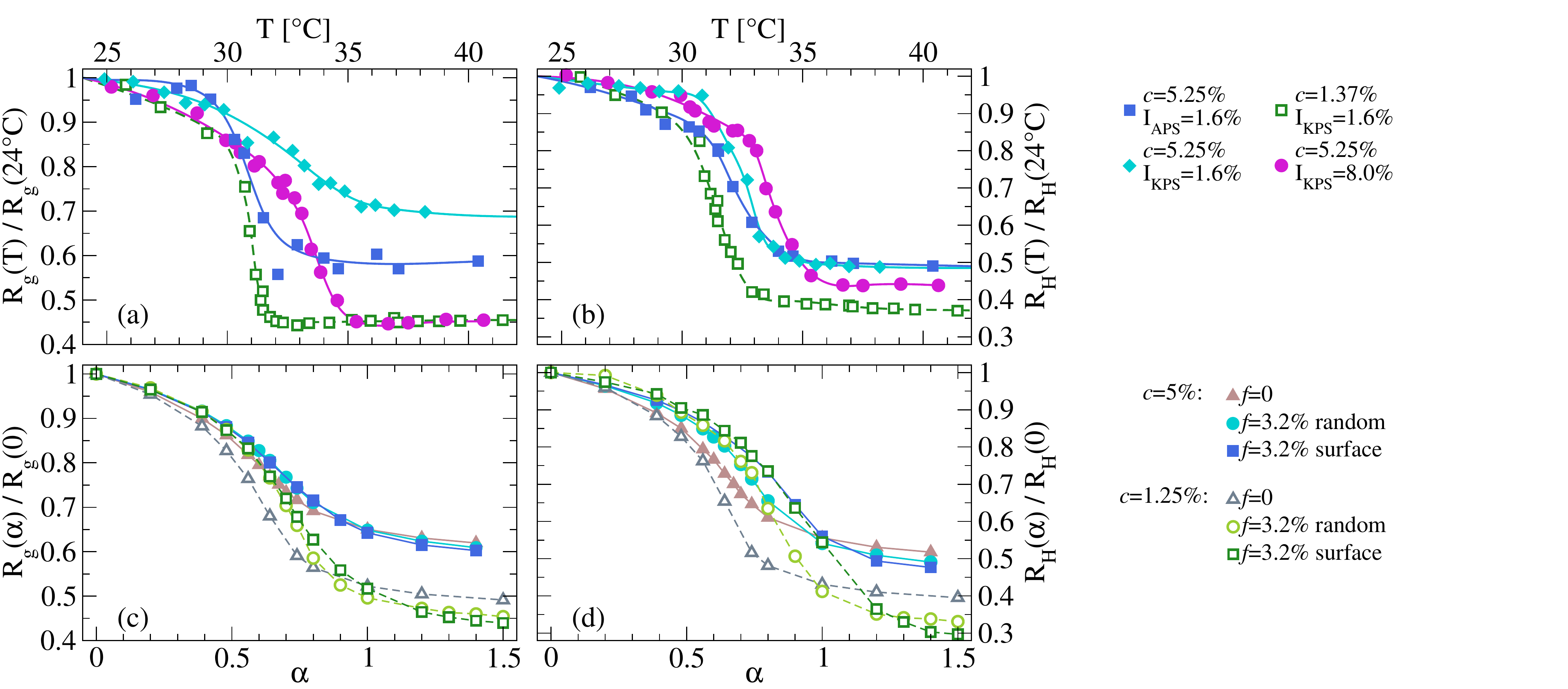}
	\caption{\label{fig:normalized}
		\textbf{Normalized swelling curves} for gyration and hydrodynamic radius for experiments:  (a) $R_g(T) /R_g(T=25$ \textcelsius), (b) $R_H(T) /R_H(T=25 $\textcelsius)  and simulations: (c) $R_g(\alpha) /R_g(\alpha=0) $, (b) $R_H(\alpha) /R_H(\alpha=0)$. Data are the same as in Fig.~\ref{fig1-2} of the main text but scaled by the respective low temperature (low $\alpha$) values. 	}
\end{figure}

In Figure~\ref{fig:normalized} we report the normalized version of the swelling curves for both \rg and \rh  from experiments and simulations, reported in Fig.~\ref{fig1-2} of the main text. This highlights the similar swelling ratios between low and high temperature obtained for experimental samples and numerical calculations, reinforcing our confidence in the employed method to estimate $R_H$ in simulations.

In addition, we discuss the transition temperatures $T_c$ of both \rg and \rh for the experimental samples, and the analogous effective temperatures $\alpha_c$ for the simulated microgels. For numerical data we simply calculate $\alpha_c$ as the inflection point of a spline fit of the data. In experiments it is more convenient to extract $T_c$ from fits of the swelling curves through the phenomenological function
\begin{eqnarray}\label{eq:swelling}
R(T) = R_0 - \Delta R\tanh{[s(T-T_c)]} +  A(T-T_c) + B(T-T_c)^2 + C(T-T_c)^3
\end{eqnarray}
that is able to accurately capture the behavior of both \rg and \rh in the whole investigated temperature range, avoiding spurious effects from the large fluctuations. We do not rely on the Flory-Rehner equation, that involves several free parameters, and moreover it has been argued not to be able to quantitatively describe the experimental swelling of PNIPAM microgels \cite{lopez2017flory}.
In Tab.~\ref{tab:1} the experimental transition temperatures for \rg and \rh are reported.

\begin{table}[h!]
	\centering
	\begin{tabular}{|c|c||c|c||c|c||c|c||}
		\hline
		\hline
		\multicolumn{2}{c}{}	&	\multicolumn{2}{c}{Fits with Eq.\ref{eq:swelling}}	&	\multicolumn{2}{c}{Fit errors}	\\
		\hline
		\hline
		& & & & & \\[-1em]
		$c$ [\%mol]	&	$I_X$ [\%mol]	&	$T_c^{R_g}$ [\textcelsius]	&	$T_c^{R_H}$ [\textcelsius]	&	$\Delta T_c^{R_g}$ [\textcelsius]	&	$\Delta T_c^{R_H}$ [\textcelsius]	\\
		\hline
		\hline
		5.25	&	1.6 (KPS)	&	31.6	&	32.4	&	0.5	&	0.8	\\
		\hline
		5.25	&	1.6 (APS)	&	30.9	&	32.1	&	1.1	&	3.4	\\
		\hline
		5.25	&	8.0 (KPS)	&	33.3	&	33.7	&	0.4	&	2.4	\\
		\hline
		1.37	&	1.6 (KPS)	&	30.9	&	31.3	&	0.3	&	1.8	\\
		\hline
		\hline
	\end{tabular}
	\vspace{0.5cm}
	\caption{\label{tab:1} \textbf{Estimated values of the VPT temperature $T_c$} obtained through fits of the swelling curves for $R_g$ and $R_H$ with Eq.~\ref{eq:swelling} (columns 3 and 4) and related errors (columns 5 and 6) for the different microgel samples. Here, $c$ and $I_X$ ($X=KPS$, $APS$) are the molar fractions used for, respectively, the crosslinker and the initiator during the synthesis.
	}
\end{table}

First of all we notice that the value of $T_c$ is similar for all three samples with $c=5.25\%$ of crosslinker, as expected. In addition, we find that the VPT temperature is slightly lower for $R_g$ with respect to $R_H$ by about $0.5\div1.0$\textcelsius. For the highly charged sample, with a nominal initiator concentration of $I_{KPS}=8.0$\si{\mole\%}, we find a larger $T_c$ for both \rg and \rh, in agreement with previous studies on charged microgels~\cite{del2020charge,capriles2008coupled}.

This scenario is in qualitative agreement with simulations data, for which the values of $\alpha_c$ are summarised in Table~\ref{tab:2}. \begin{table}[h!]
	\centering
	\subfloat[neutral microgels]{
		\begin{tabular}{|c|c||c|c|}
			\hline
			\hline
			$c$ [\%]	&	$f$ [\%]	&	$\alpha_c^{R_g}$	&	$\alpha_c^{R_H}$	\\
			\hline
			\hline
			1.25	&	0	&	0.62	&	0.65	\\
			\hline
			3.0	&	0	&	0.64	&	0.66	\\
			\hline
			5.0	&	0	&	0.68	&	0.69	\\
			\hline
			10.0	&	0	&	0.60	&	0.65	\\
			\hline
			\hline
		\end{tabular}
	}
	\subfloat[random charge distributions]{
		\begin{tabular}{|c|c||c|c|}
			\hline
			\hline
			$c$ [\%]	&	$f$ [\%]	&	$\alpha_c^{R_g}$	&	$\alpha_c^{R_H}$	\\
			\hline
			\hline
			1.25	&	1.0	&	0.66	&	0.69	\\
			\hline
			3.0	&	1.0	&	0.63	&	0.69	\\
			\hline
			5.0	&	1.0	&	0.66	&	0.68	\\
			\hline
			1.25	&	2.0	&	0.70	&	0.77	\\
			\hline
			1.25	&	3.2	&	0.76	&	0.78	\\
			\hline
			3.0	&	3.2	&	0.73	&	0.82	\\
			\hline
			5.0	&	3.2	&	0.69	&	0.75	\\
			\hline
			10.0	&	3.2	&	0.69	&	0.72	\\
			\hline
			3.0	&	10	&	0.89	&	1.04	\\
			\hline
			5.0	&	10	&	0.79	&	1.04	\\
			\hline
			\hline
		\end{tabular}
	}
	\subfloat[surface charge distributions]{
		\begin{tabular}{|c|c||c|c|}
			\hline
			\hline
			$c$ [\%]	&	$f$ [\%]	&	$\alpha_c^{R_g}$	&	$\alpha_c^{R_H}$	\\
			\hline
			\hline
			1.25	&	3.2	&	0.72	&	0.87	\\
			\hline
			3.0	&	3.2	&	0.69	&	0.95	\\
			\hline
			5.0	&	3.2	&	0.63	&	0.90	\\
			\hline
			10.0	&	3.2	&	0.77	&	0.86	\\
			\hline
			5.0	&	10	&	0.66	&	-	\\
			\hline
			\hline
		\end{tabular}
	}
	\subfloat[charges on surface chains ($f=3.2\%$)]{
		\begin{tabular}{|c|c||c|c|}
			\hline
			\hline
			$c$ [\%]	&	$q*$ per chain [$-e*$]	&	$\alpha_c^{R_g}$	&	$\alpha_c^{R_H}$	\\
			\hline
			\hline
			5.0	&	1	&	0.72	&	0.72	\\
			\hline
			5.0	&	2	&	0.68	&	0.78	\\
			\hline
			5.0	&	3	&	0.68	&	0.87	\\
			\hline
			\hline
		\end{tabular}
	}
	\caption{\label{tab:2}  {\textbf{ Estimated values of the solvophobic parameter $\alpha_c$ at the VPT }} obtained through splined fits of the swelling curves for $R_g$ and $R_H$ for numerical microgels. Here, $c$ and $f$ are the crosslinker and the charged beads percentage, respectively. The fit error is negligible compared to the discretisation error, for which an upper limit can be quantified as $\Delta\alpha_c \sim 0.05$.
	}
\end{table}
Similarly to what observed in experiments, the value of $\alpha_c$ is found to depend on the amount of charges on the microgels, while the dependence on crosslinker concentration seems to be negligible. Indeed, for a given value of $f$, differences in the data are found to be within the error, considering this to be of the order of 0.05, 
given the used sampling of $\alpha$.
In particular, we find that it increases with increasing $f$ and also moving from a random to a surface distribution of the ionic monomers.
Again, we also find that the estimated $\alpha_c$ is slightly larger when estimated from $R_H$ than from $R_g$. We notice that for a microgel  with c=5\% and f=10\% arranged on the surface, a robust value of $\alpha_c$ could not be estimated because we did not detect the full collapse of the microgel.

\section*{Electrophoretic mobility}
\begin{figure}[!h]
	\centering
	\includegraphics[width=9cm]{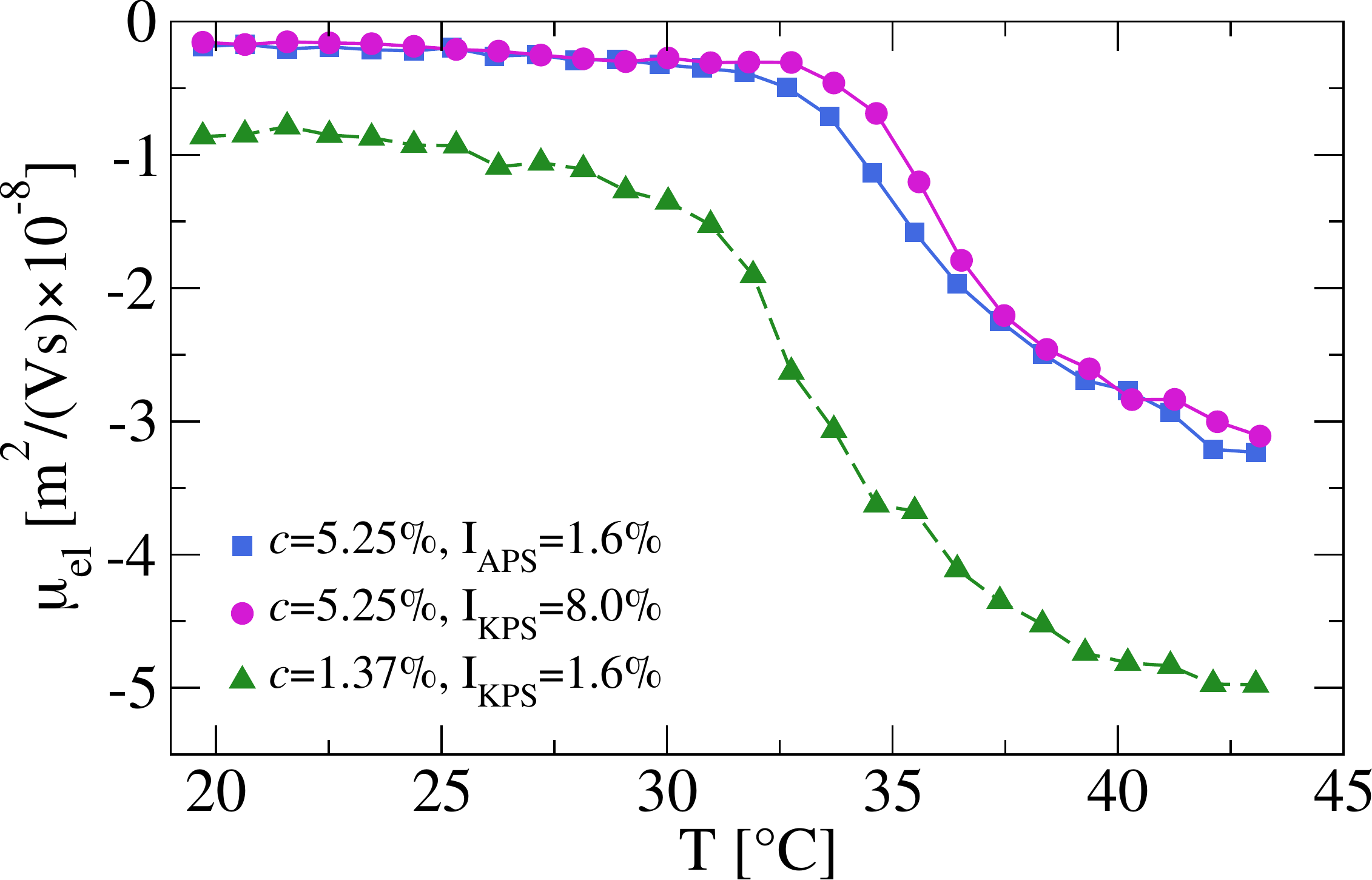}
	\caption{\label{fig:mobility}
		\textbf{Electrophoretic mobility} measurements as a function of temperature $T$ on samples with different size, charge and crosslinker content. The relative error on measurements is less than $5\%$ (not shown).	}
\end{figure}

In Fig.~\ref{fig:mobility} we show the $T$-dependence of the electrophoretic mobility \mel measured for three of the samples discussed in the manuscript. In agreement with previous works~\cite{truzzolillo2018overcharging}, we find that a strong increase of \mel is observed close to the VPT transition.  In particular, we observe that c=5\% microgels (squares and circles in Fig.~\ref{fig:mobility}) at low $T$ have much lower values of \mel with respect to c=1.37\% microgels (triangles in Fig.~\ref{fig:mobility}) at the same charge fraction. This may be due to the larger size of the former, since in the swollen state they have lower diffusivity and better screening conditions (lower surface-to-volume ratio).
The transition temperature seems to increase with increasing initiator concentration while not being affected from the crosslinker concentration.
Surprisingly, microgels with different initiator quantities but of roughly the same size show quite similar values of \mel. This may be due to different factors: on one hand, we expect that, for large $I_{KPS}$, a non-negligible part of the initiator comes out in low molecular weight byproducts, given the high reactivity of these molecules, resulting in a bare charge for final microgels particles which is lower than the nominal one; on the other hand, the higher the bare charge, the stronger is the electrostatic screening, resulting in an effective charge which is not proportional to the bare one, if not for very low values. Further experiments assessing the dependence of \mel on initiator content in a more systematic fashion would be needed in order to properly understand this behavior, which is beyond the scope of the present work.

\section*{On the estimate of \rh}
As reported in the main text, the hydrodynamic radius \rh is a measure of the diffusivity of the particles, that is defined as the radius of a spherical particle moving in a continuous viscous fluid with small Reynolds number (Stokes radius):
\begin{equation}
R_H = \frac{k_B T}{6\pi\eta \bar{D}}
\end{equation}
where $\bar{D}$ is the long-time diffusion coefficient, $k_BT$ the thermal energy, and $\eta$ the dynamic viscosity.
For suspensions of Hard Spheres (HS) \rh roughly corresponds to the geometrical radii of the particles, while for polymeric soft particles, for which a sharp geometrical surface cannot be defined, it still depends on the extension of the external polymeric chains and their density.
Since in our simulations with implicit solvent treatment the diffusion coefficient cannot be computed, and simulations in explicit solvent (with even coarse-grained particles) are out of our disposal of computing resources, we are left to estimate \rh from static equilibrium properties.
\begin{figure}[b!]
	\centering
	\includegraphics[scale=0.4]{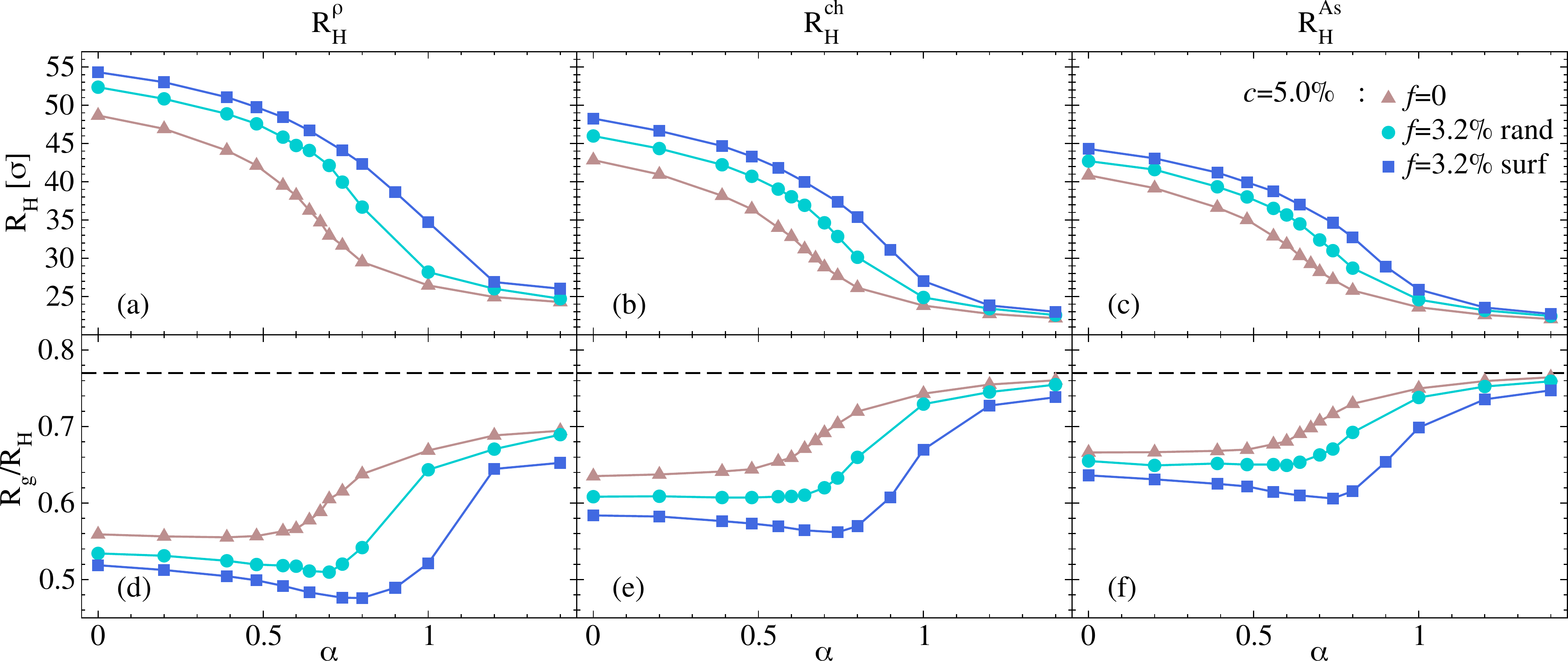}
	\caption{\label{fig:rh_versions}
		\textbf{Comparison among different estimates of \rh.} Top panels show the hydrodynamic radius as a function of the solvophobic parameter $\alpha$ for microgels with c=5.0\%.
		According to the methods described in the text, the different panels report: $R_H^\rho$ (a), $R_H^{ch}$ (b) and $R_H^{As}$ (c) for neutral and charged microgels ($f=3.2\%$) with random and surface distribution. Bottom panels show the corresponding values of $R_g/R_H$. The surface mesh used to calculate $R_H^{As}$ (c,f) has been constructed using a probing radius of $12\sigma$. }
\end{figure}
Several approximated methods have been proposed to compute estimated values of \rh from equilibrium configurations of polymeric particles, each with different limitations\cite{mansfield2007comparison}. All of them consider the solvent as a continuous medium flowing through and around the particles.
We did not consider methods based on the approximation of the macromolecule as a continuous porous medium\cite{brinkman1949calculation,ooms1970frictional}, since this approximation is not suitable for polymeric particles with low chains density in the corona, such as microgels.
We also avoided the use of the Kirkwood double-sum formula\cite{kirkwood1996general}, which does not account for long range correlations in the hydrodynamic force field, and it has been applied successfully only to simple polymer chains.
The most accurate methods for estimating the hydrodynamic radius are based on numerical or approximated steady-state solutions of the Navier-Stokes equations for ensembles of spherical bead scatterers\cite{swan2016rapid,ortega2011prediction}, which are highly computationally demanding for large sized macromolecules.
In any case, for systems such as microgels, the main effect on the diffusivity of the particles is given by the polymer chains on the corona, hence it is convenient to define a particle surface\cite{mansfield2007comparison} and then estimate \rh as the hydrodynamic radius associated to that shape. In this work we pursue that way, which represents a trade-off among accuracy and computational feasibility.

In this work,  we instantaneously approximate the microgel to an effective ellipsoid~\cite{rovigatti2019elasticity} and then calculate the hydrodynamic friction using the approach developed by Hubbard and Douglas~\cite{hubbard1993hydrodynamic}. For each instantaneous configuration, we first compute the convex hull containing all the beads of the microgel and then evaluate the gyration tensor of all the simplices it is made of. Finally, we determine the ellipsoid having the same gyration tensor as previously done in Ref.~\cite{rovigatti2019elasticity}.
To compute the hydrodynamic friction $\zeta$ of each ellipsoid we exploited the relation~\cite{hubbard1993hydrodynamic}:
\begin{equation}
\zeta = 6\pi\eta C_\Omega \equiv 6\pi\eta R_H
\label{eq:friction}
\end{equation}
where $\eta$ is the solvent viscosity and $C_\Omega$ is the electrostatic capacitance. This relation has been shown to well approximate the friction acting on brownian rigid particles and to be valid for objects of arbitrary shape~\cite{douglas1994hydrodynamic}.
Eqn.~\ref{eq:friction} is based on a solution of the Navier-Stokes equation for steady flow of rigid particles with stick boundary conditions, where the hydrodynamic interactions are described by the isotropic angular averaged Oseen tensor~\cite{hubbard1993hydrodynamic}. In this approximation, the Navier-Stokes equation for the momentum flux density assumes the same form of the Poisson's equation for electrostatics, thereby the hydrodynamic radius becomes mathematically equivalent to the electrostatic capacitance, which for rigid ellipsoids is~\cite{hubbard1993hydrodynamic}:
\begin{equation}
C_\Omega = 2\left[ \int_{0}^{\infty}\frac{1}{\sqrt{(a^2+\theta)(b^2+\theta)(c^2+\theta)}}d\theta \right]^{-1}
\label{eq:capacitance}
\end{equation}
where $a$, $b$, $c$ are the principal semiaxes.

Here we show the comparison among three methods to compute \rh by defining an effective surface for the microgel.
For the first one we approximate the microgel as a sphere of radius $R_H^\rho$ such that the equilibrium monomers density profile assumes the value $\rho(R_H^\rho) = 10^{-3}\sigma^{-3}$ at this distance from the center of mass, as also done in a previous work \cite{ninarello2019modeling}. This definition results to be rough as compared to the other two methods, because it does not account for thermal and shape fluctuations.
For the other two methods we compute for each equilibrium configuration a surface mesh, approximating the microgel as a hard ellipsoid having the same gyration tensor of the surface faces, of which we compute the hydrodynamic radius as explained in the main text. The difference among them consists in the way the surface mesh is constructed. In one case the convex hull has been computed, obtaining the estimated hydrodynamic radius $R_H^{ch}$, which is the value used in the computations of the main text. The other case involves the computation of the surface mesh with the Alpha-shape method\cite{stukowski2014computational}, which is based on a Delaunay tessellation of the microgels' beads, thereby the space regions are considered as empty or occupied whether the radius of a circumscribing sphere is greater or smaller than a certain probing radius $r_p$, after that a surface mesh is constructed. At the same way, from this surface mesh (computed exploiting the Ovito python package) is obtained the hydrodynamic radius estimate $R_H^{As}$, which obviously depends on the probing sphere radius $r_p$. The last two methods lead to comparable results, but the convex hull method is parameter-free and the values of the ratio $R_g/R_H^{ch}$ at small and large values of the solvophobic parameter $\alpha$ seems to be more in agreement with the experimental observations, hence we decided to use that method throughout the manuscript.

In Fig.~\ref{fig:rh_versions} we show the comparison among the three methods by observing the different estimates of \rh (top panels) and the ratio $R_g/R_H$ (bottom panels) on microgels with crosslinker concentration $c=5.0\%$ (similar trends are observed for microgels with different values of $c$, not shown), different charged beads fractions ($f=0, 3.2\%$) and charge distributions (random, surface).
The calculation of $R_H^\rho$ is found to be less accurate with respect to the other two methods, due to the low statistics of monomers for densities $\rho(r)\sim 10^{-3}\sigma^{-3}$, well in the outer corona region. In addition, the other two methods both tend to the correct HS value for \ratio at large $\alpha$.
Importantly, all three methods give an overall qualitatively similar behavior, with the onset of a minimum in $R_g/R_H$ for microgel with surface arrangement of charges, while the neutral microgel does not show such a minimum for all three definitions of hydrodynamic radius.

\begin{figure}[!t]
	\centering
	\includegraphics[width=7.5cm]{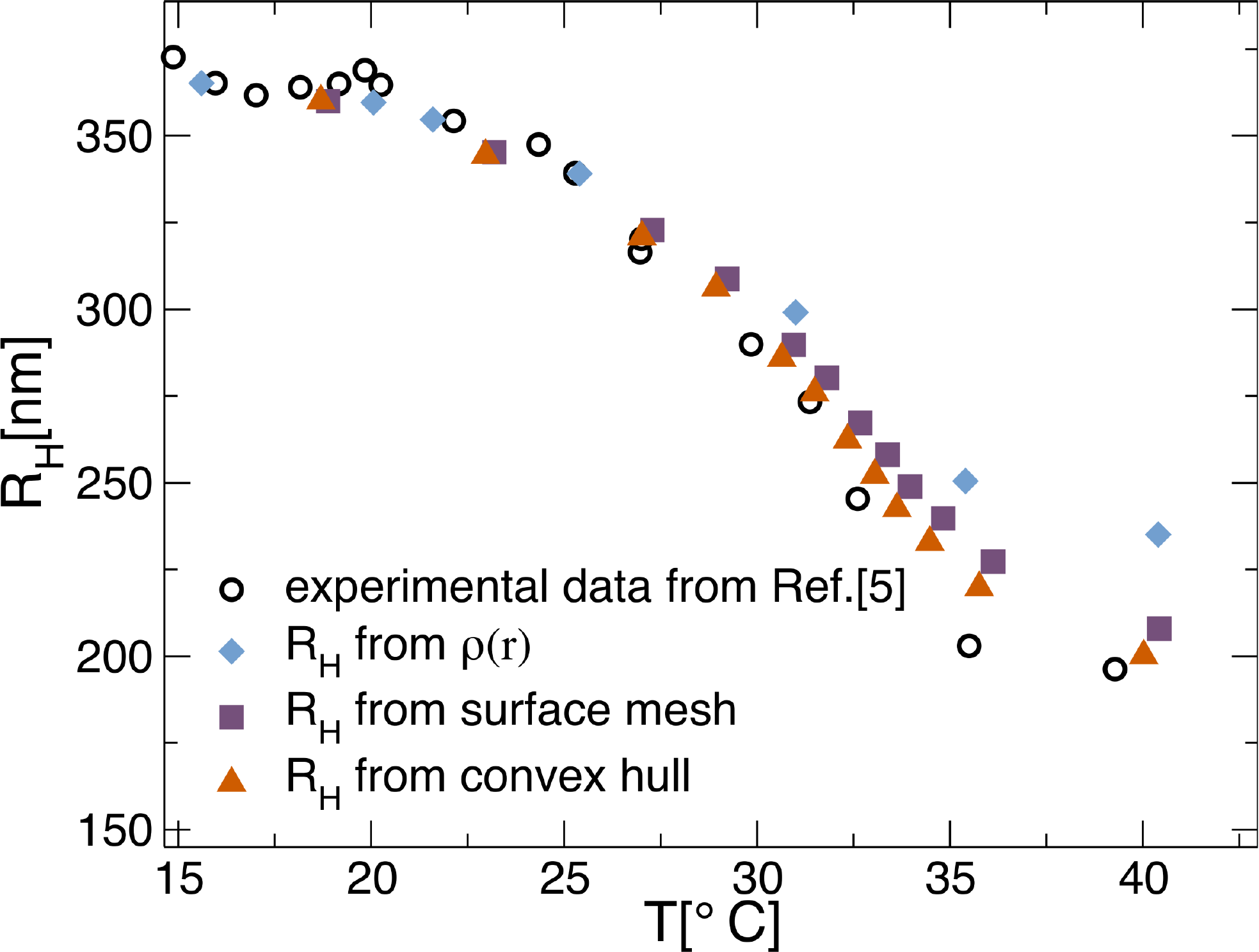}
	\caption{\label{jerome}
		\textbf{Comparison between $R_H$ from experiments and simulations:} DLS data for the hydrodynamic radius are taken from Ref.~\protect\cite{ninarello2019modeling} and compared with the estimate based on the threshold of the monomer density profiles (diamonds), the one from the surface mesh with probe radius equal to 12$\sigma$ (squares) and the present one based on the convex hull (triangles). Both simulation data are for N=42K neutral microgels and we used the same $\alpha-T$ relationship as in previous work, without additional fit parameters.}
\end{figure}

To further validate our current definition of $R_H$, we here perform a quantitative comparison with experimental data reported in Ref.~\cite{ninarello2019modeling}. In this case, a very large amount of salt was added to the samples in order to screen electrostatic effect, so that the comparison can be performed using neutral microgels. Fig.~\ref{jerome} reports $R_H$ measured by DLS as a function of $T$ that is compared to the hydrodynamic radius estimated in simulations within the three approaches described above.  The comparison is performed using the experimental-numerical mapping established in Ref.~\cite{ninarello2019modeling} based on the $T$-dependence of the form factors, i.e. without employing additional fit parameters.
We find a significant improvement of the agreemenent between experiments and simulations when adopting the present definition of $R_H$, particularly with respect to the one based on $\rho(r)$. Indeed, the full collapse of the microgel appears to be well reproduced by the simulations. These findings provide additional robustness to the present method used to define $R_H$. However, it will be important in the future to perform a similar kind of comparison with experiments, based on the form factors, also for charged microgels.

\section*{Local swelling in monomers profiles}
\begin{figure}[h!]
	\centering
	\includegraphics[width=15.cm]{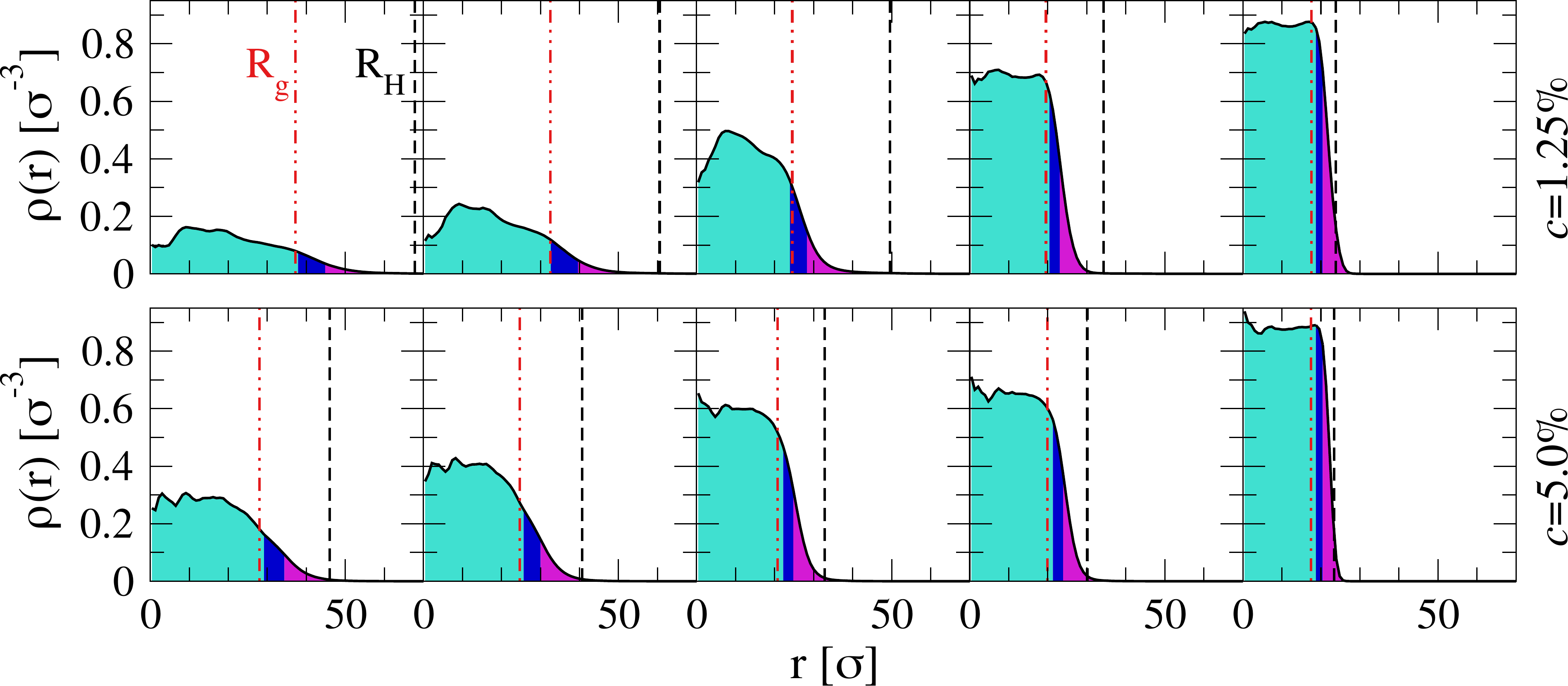}
	\caption{\label{fig:local-profiles}
		\textbf{Local swelling profiles.} Monomer density profiles for a microgel with $c=1.25\%$ and $f=3.2\%$ randomly distributed charges (top panels) for effective temperature values $\alpha = 0, 0.48, 0.74, 0.90, 1.20$, and for a a microgel with $c=5.0\%$ and $f=3.2\%$ randomly distributed charges (bottom panels) for effective temperature values $\alpha = 0, 0.48, 0.74, 0.80, 1.20$. The different colors indicate the monomers corresponding to the three regions considered for the analysis of the local swelling, region I (cyan), region II (blue) and region III (magenta). Vertical lines indicate \rg (red, dash-dotted) and \rh (black dashed) values.
	}
\end{figure}
In Fig.~\ref{fig:local-profiles} we show the monomer density profiles calculated from simulations of microgels with $c=1.25\%, 5.0\%$ (top and bottom panels, respectively) and $f=3.2\%$ randomly distributed charges as a function of $\alpha$.
The coloured regions beneath the curves represent the three different regions used in the analysis of the local swelling of Fig.~\ref{fig:local_swelling} of the main text, highlighting the core (region I, containing 65\% of most inner particles), the inner corona (region II, consisting of 20\% of monomers nearest to the core beads) and the outer corona (region III, comprising the 15\% most exterior beads)  of the microgels. 
Representative snapshots corresponding to microgels in each panel of the figure are shown in Fig.~\ref{fig:snaps} of the main text.
As it can be seen, the gyration radius \rg roughly corresponds to the extension of the core region at all $c$ and $\alpha$ values, sharing similar swelling properties as evident from Fig.~\ref{fig:local_swelling} of the main text, while the hydrodynamic radius \rh embeds also the corona regions, and its value is clearly dependent on the extension of the outer chains, with differences among the cases $c=1.25\%$ and $c=5.0\%$.
For completeness, we notice that the nonmonotonicities observable at low-distance are due to the fact that these data are not averaged over multiple initial configurations. They would be removed by considering several microgel topologies, as shown in our previous works~\cite{ninarello2019modeling}, and do not affect the local swelling discussion.

\section*{Chain length distributions}
\begin{figure}[!h]
	\centering
	\includegraphics[scale=0.4]{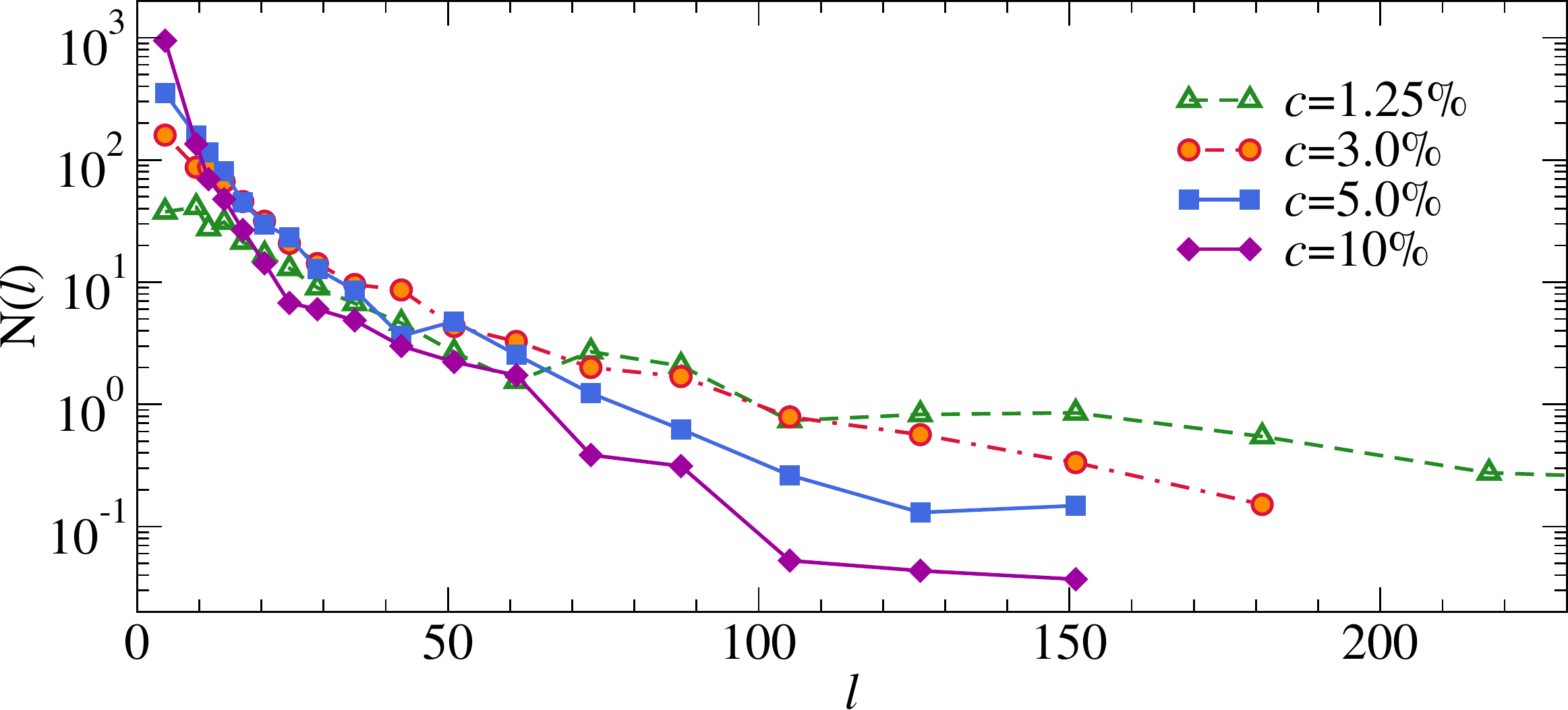}
	\caption{\label{fig:chain-lengths}
		\textbf{Chain length distributions} for microgels with different crosslinker concentrations $c$.}
\end{figure}

Fig.~\ref{fig:chain-lengths} shows the number distribution of the chain length $l$ of the microgels. The behavior of all the curves is consistent with a superposition of two exponential distributions, one for low $l$ values, mainly accounting for the chains in the core region, and another for high values of $l$, which is mostly given by the long chains that can be found in the corona of the microgels. This is due to the radial distribution of the crosslinkers used in the assembly process of the network topology \cite{ninarello2019modeling}, with the aim to realistically mimic the mass distribution of microgels. Of course, it is evident that as $c$ decreases, small chains become less and less populated while longer chains  increase in number and also in length. Again, the noise of the data at large $l$ would be easily improved by considering averages over multiple microgel topologies, as already done in Ref.~\cite{ninarello2019modeling}.

\section*{Local swelling for surface charged microgels}

\begin{figure}[h!]
	\centering
	\includegraphics[width=15.cm]{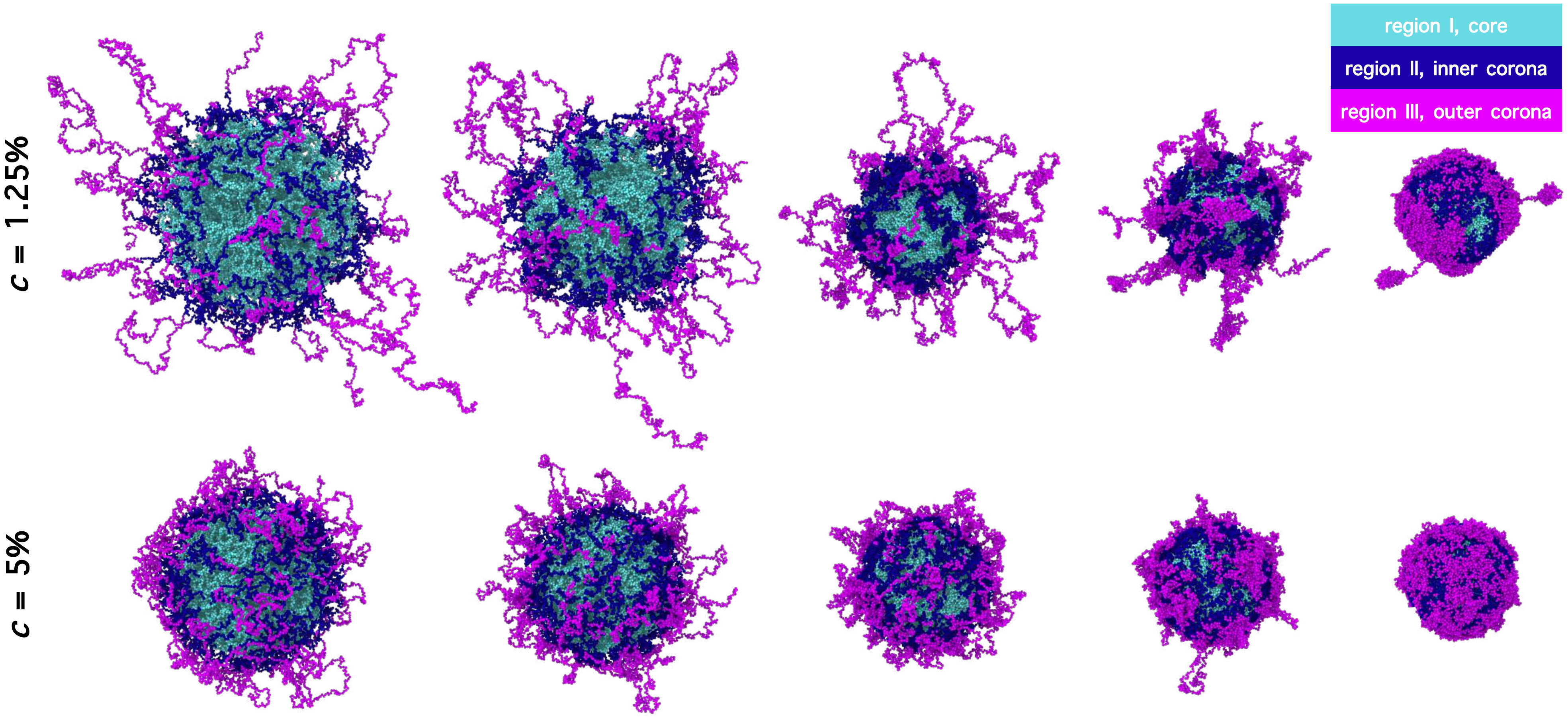}
	\caption{\label{fig:surfchargedmgel}
		\textbf{Snapshots of surface charged microgels across the VPT illustrating the local swelling.} Simulation snapshots for a surface-charged microgel ($f=3.2\%$) for different values of the swelling parameter $\alpha$ from the swollen to the collapsed state. From left to right, for $c=1.25\%$, $\alpha=0.0, 0.48, 0.80, 0.90, 1.20$, while for $c=5\%$, $\alpha=0.0, 0.48, 0.74, 0.90, 1.20$. Monomers are colored according to the region they belong to: cyan indicate the core region (defined as 65\% of most inner particles), blue the inner corona region (consisting of 20\% of monomers nearest to the core beads) and purple the outer corona region (comprising the 15\% most exterior beads). 
	}
\end{figure}

For the sake of completeness, we report in Fig.~\ref{fig:surfchargedmgel} the snapshots showing the local swelling of surface-charged microgels for $c=1.25\%$ and  $c=5\%$ with a charge fraction $f=0.032$.

\section*{Effects of charge fraction for randomly charged microgels}
\begin{figure}[h!]
	\centering
	\includegraphics[scale=0.4]{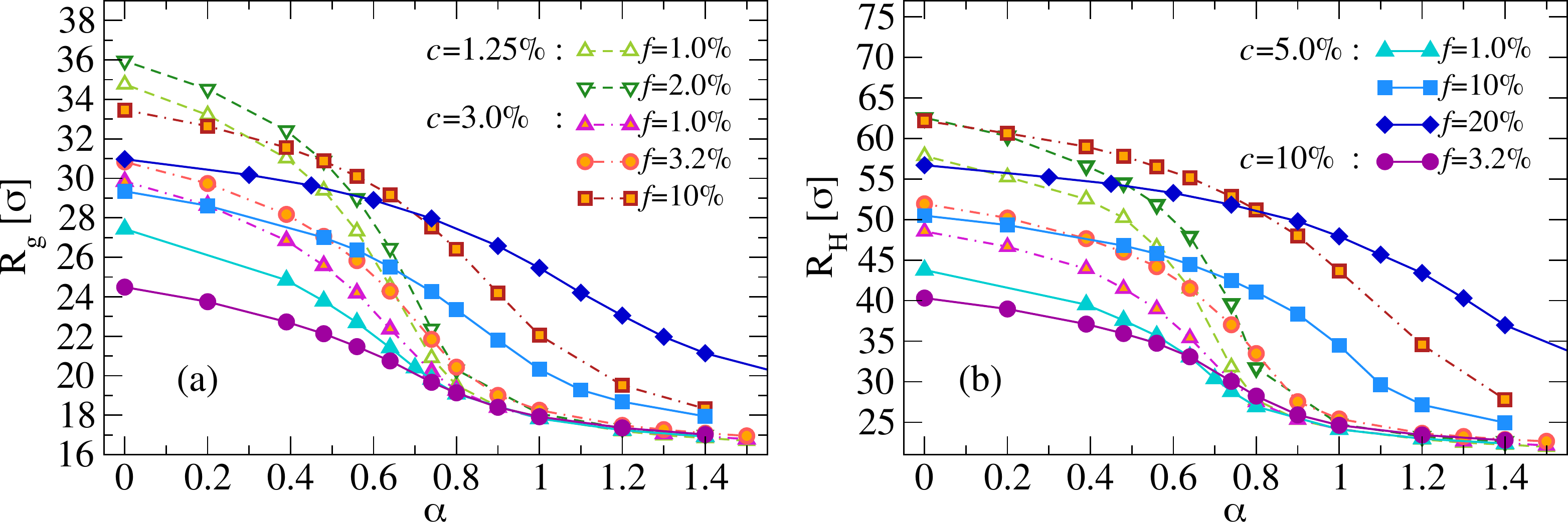}
	\caption{\label{fig:rg-rh-random}
		\textbf{Swelling curves for randomly charged microgels}. \rg and \rh swelling curves for randomly charged microgels with different crosslinker ($c$) and charge ($f$) content. }
\end{figure}

In Fig.~\ref{fig:rg-rh-random} we show the swelling curves of \rg and \rh for randomly charged microgels, whose ratio \ratio is reported in Fig.~\ref{fig:random}(a) of the main text. The swollen size of microgels and the transition temperature $\alpha_c$ increase as $f$ increases, as shown in previous studies. The swelling ratio, defined as the ratio between the swollen and the collapsed radius, is found to be inversely proportional to $c$, as expected.

\section*{Charged monomers distributions in microgels with charges on the surface}
\begin{figure}[h!]
	\centering
	\includegraphics[scale=0.4]{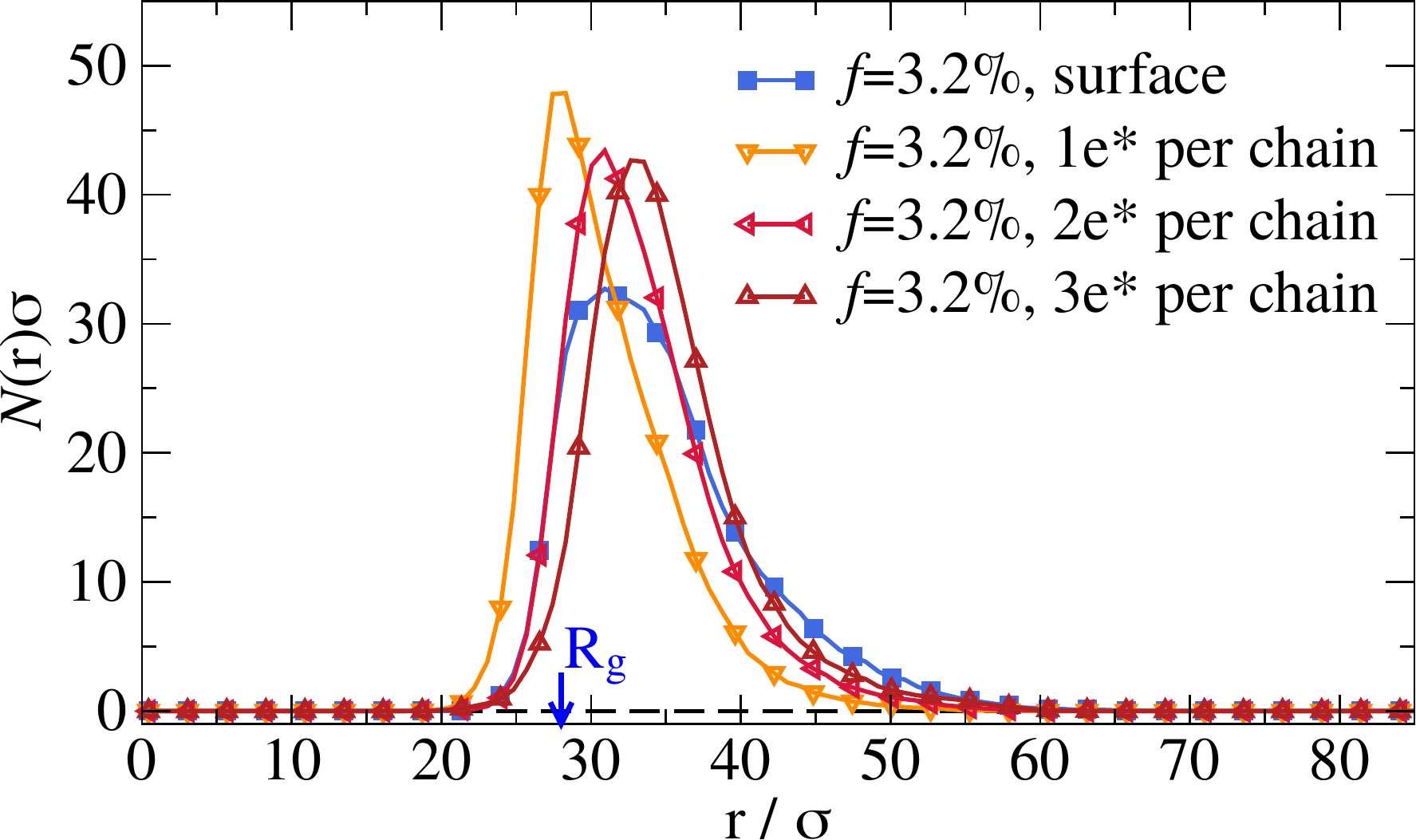}
	\caption{\label{fig:ions-profs-surface}
		\textbf{Charged monomers radial distributions} of surface charged microgels with crosslinker concentration $c=5.0\%$ and charges fraction $f=3.2\%$. Different ways of distributing charged beads on the microgels' surface are considered. }
\end{figure}

In Fig.~\ref{fig:ions-profs-surface} we show the radial distribution of charged monomers in the four kinds of surface charges distributions with $f=3.2\%$ displayed in Fig.~\ref{fig:surface} of the main text.

\section*{Effect of charge distribution}
\begin{figure}[h!]
	\centering
	\includegraphics[height=3.75cm]{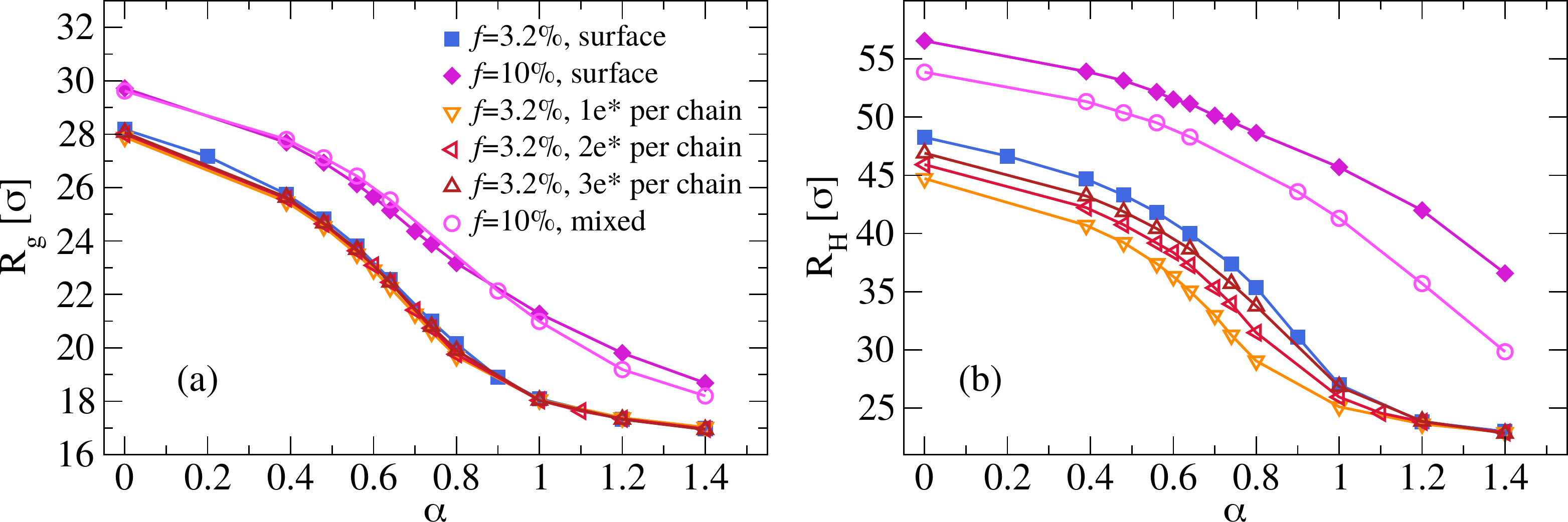}
	\includegraphics[height=3.75cm]{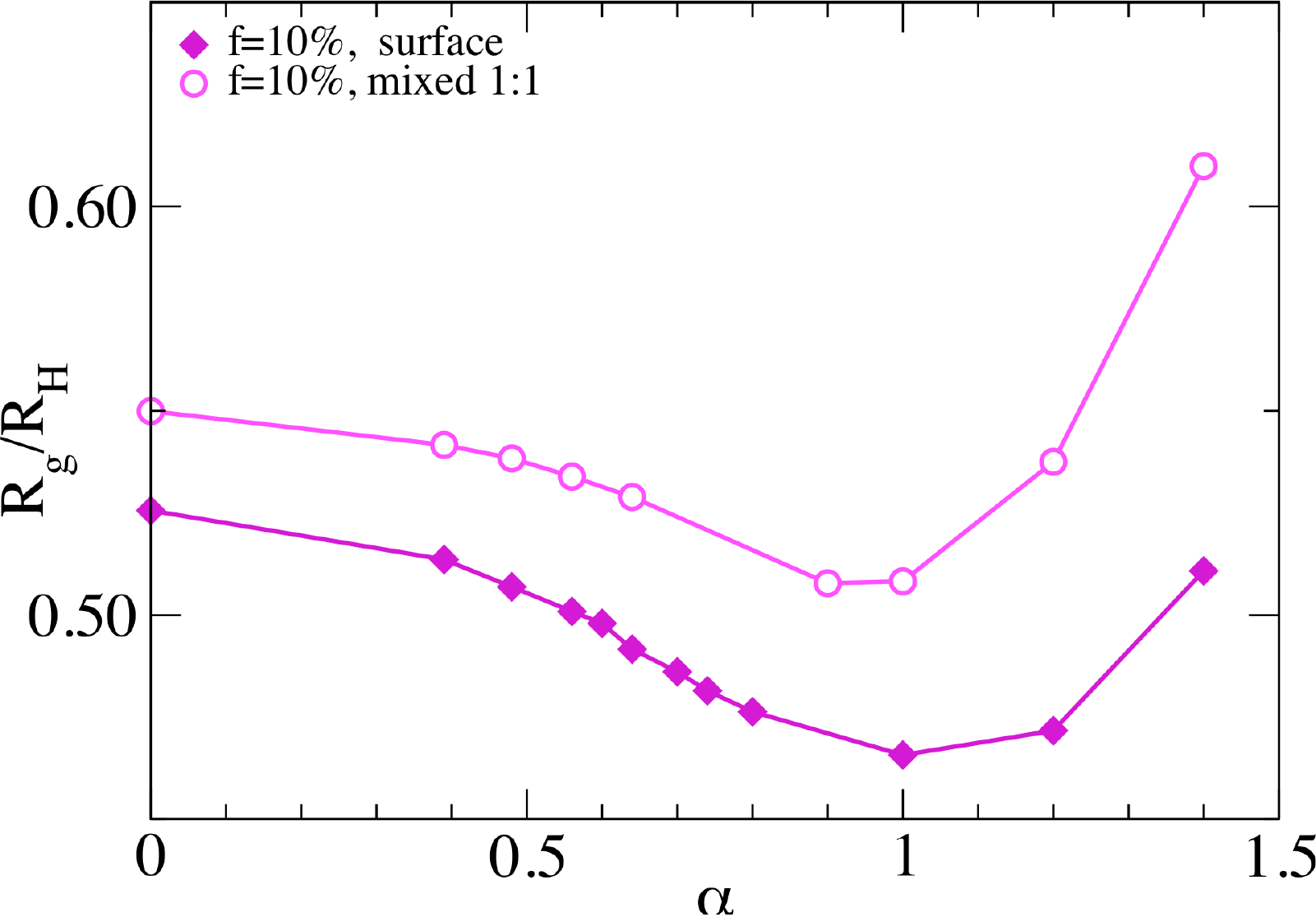}
	\caption{\label{fig:rg-rh-surface}
		\textbf{Swelling curves of surface charged microgels}. \rg and \rh swelling curves for surface charged microgels with crosslinker concentration $c=5.0\%$ and charges fractions $f=3.2\%, 10\%$. Different ways of distributing charged beads on the microgels' surface are considered. Also the ratio $R_g/R_H$ is reported for high charge content ($f=10\%$).}
\end{figure}

\begin{figure}[t!]
	\centering
	\includegraphics[scale=0.4]{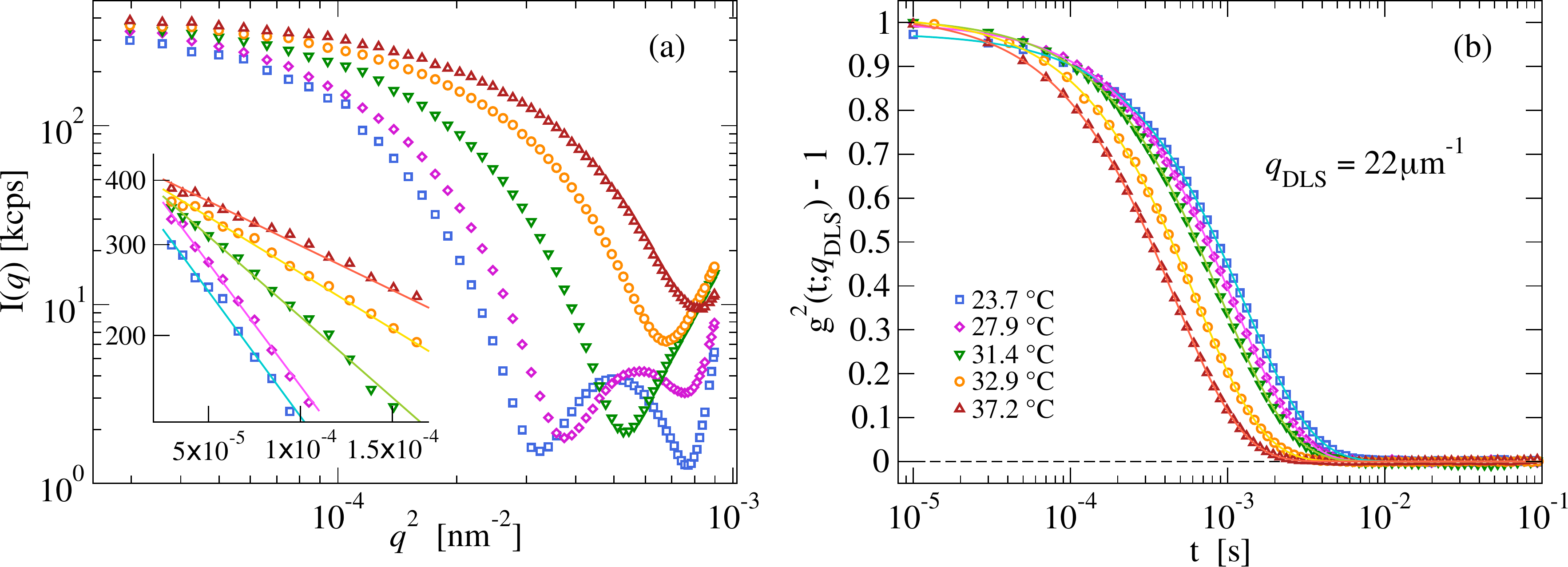}
	\caption{\label{fig:Ik-gtk}
		\textbf{SLS and DLS measurements.} (a) Intensity I(q) scattered by a sample of microgels with $c = 5.25$\si{\mole\%} and $I_\text{APS} = 1.6$\si{\mole\%} prepared as detailed in the main text at different temperatures across the VPT.
		Here $q = 4\pi n_0 \sin(\theta/2)/\lambda$ is the magnitude of the scattering vector, $\theta$ the scattering angle, $\lambda=532$\si{\nano\meter} is the wavelength of the laser beam, and $n_0=1.33$ the refractive index of the sample. The inset shows the low-$q$ part of the scattering intensity, where the Guinier regime is attained. The solid lines are non-linear regressions obtained via Eq.~1 of the main text.
		(b) Intensity autocorrelation functions $g^2(t;q_\text{DLS})-1$, proportional to the intermediate scattering functions $F_s(q_\text{DLS},t)$, measured at $q_\text{DLS}=22$\si{\micro\meter^{-1}}; solid lines are the best fits obtained via Eq.~3 of the main text. Temperatures for both datasets are indicated in panel (b).
	}
\end{figure}

Fig.~\ref{fig:rg-rh-surface} shows the swelling curves of \rg and \rh for surface charged microgels whose ratio \ratio is displayed in Fig.~\ref{fig:surface} of the main text.
Different ways to distribute charged beads on the microgels' corona are considered. We report the case of standard surface-charged microgels, for which charges are randomly distributed in the region $r>R_g$ on a neutral equilibrated configuration with the constraint that two consecutive beads on a chain cannot both be charged. In addition, we consider $f=3.2\%$ microgels in which charges are distributed in a constant number for each chain on the most exterior chains. As we can notice, \rg is almost not affected by the charge distribution, while \rh significantly increases as the number of charges per chain grows. We also consider, for the highly charged microgel ($f=10\%$), the case in which half of the charges are distributed randomly throughout the network, half being on the surface (mixed distribution). Also in this case the distribution has a negligible effect on the swelling properties of \rg, while having a strong effect on \rh, for which we cannot observe a transition in the range of examined $\alpha$ values.

In Fig.~\ref{fig:rg-rh-surface} we also report \ratio for charged microgels with different surface distributions, in particular focusing on high 
charge content  $f=10\%$. In this case, we do not observe a total collapse of the microgel at high $\alpha$, but still we find a rather deep minimum in \ratio for $\alpha \sim 1.0$. Such a minimum appears to be wider with respect to that occurring at low $f$ values, but the total hindering of the collapse indicates that probably such surface-charged microgel is not realizable in the experiments, given the high charge content. Hence, in addition, we analysed mixed charge distributions, still with $f=10\%$ but where half of the charges are located at random and half on the surface (mixed 1:1 distribution). These systems display an intermediate behaviour between the random and the surface charged ones.

\begin{figure}[h!]
	\centering
	\includegraphics[scale=0.4]{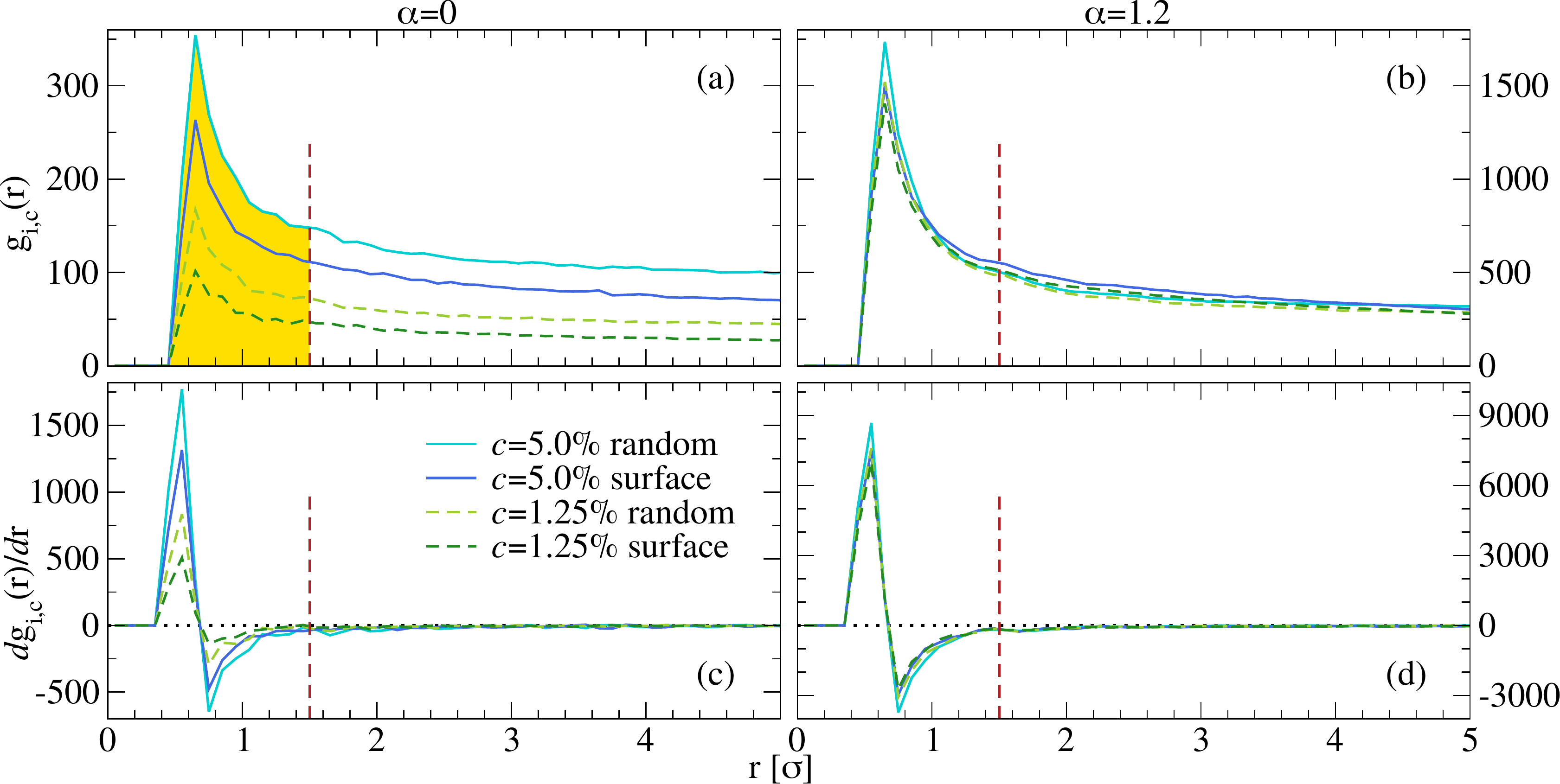}
	\caption{\label{fig:gr-ions-cions}
		\textbf{$g_{i,c}(r)$ of ions and counterions.} (a,b) cross radial distribution functions $g_{i,c}(r)$ for ions and counterions for random and surface charged microgels with $c=1.25\%, 5.0\%$ and $f=3.2\%$ at $\alpha=0$ and $\alpha=1.20$. (c,d) Derivative of $g_{i,c}(r)$ with respect to $r$ shown in the top panels. The vertical dashed line indicates the cutoff value of $r^*=1.5\sigma$ }
\end{figure}

\section*{Guinier plots and cumulants analysis}
We extracted the average gyration radius of microgels, fitting the low scattering-vector part of the intensity $I(q)$ with the functional form (Guinier approximation) as:
\begin{equation}\label{eq:guinier}
I(q) = I(0) e^{-\frac{(qR_g)^2}{3}}
\end{equation}
where $I(0)$ is a constant depending on the number of particles in the scattering volume and on the scattering factor of a single particle \cite{beaucage1995approximations}, while \rg is their radius of gyration, defined as:
\begin{equation}
R_g = \sqrt{ \frac{\sum_{i=1}^{N}(\vec{r}_i-\vec{r}_\text{cm})^2 }{ N } }
\end{equation}
and $q$ is the magnitude of the scattering vector.
The Guinier regime for our microgels is attained for $qR_g<2$ coherently with previously reported microgel syntheses~\cite{clara2012structural,gasser2014form}. For all samples Eq.~\ref{eq:guinier} reliably fitted the available data in the entire range $5.5$\si{\micro\meter^{-1}}$\le q \le 2/R_g$. The uncertainty on $R_g$ is given by the fit error, the latter being less than $10\%$ of the best-fit value.

\section*{Charges screening}
In the top panels of Fig.~\ref{fig:gr-ions-cions} the cross radial distribution functions $g_{i,c}(r)$ for ions and counterions are shown, for the swollen (a) and the collapsed (b) state. The vertical dashed line indicates the cutoff value of $r^*=1.5\sigma$, that we used to calculate the screened charge per chain used to predict the presence of the minimum in \ratio, shown in Fig.~\ref{fig:qout} of the main text. As we see from the bottom panels in Fig.~\ref{fig:gr-ions-cions}(c,d), $r^*$ roughly indicates the distance at which the derivative of $g_{i,c}(r)$ becomes roughly zero. Integrating $g_{i,c}(r)$ up to $r^*$ we estimate the amount of counterions which can be considered as ``bounded'' to charged beads, that we use to compute the effective charges $q_\text{ch}^\text{eff}$.

\providecommand{\noopsort}[1]{}\providecommand{\singleletter}[1]{#1}%

\end{document}